\documentclass[12pt,a4paper]{article}

\usepackage{feynmp}
\usepackage{geometry}
\usepackage{amsmath}
\usepackage{amssymb}
\usepackage{amsmath}
\usepackage{amsthm}
\usepackage{xspace}
\usepackage{graphicx}
\usepackage[nosort]{cite}
\usepackage{subfigure}
\usepackage{verbatim}

\newcommand{\p}{\partial_{+}}
\def\pbar{\partial_{-}}

\def\[{\begin{equation}}
\def\]{\end{equation}}
\newcommand{\be}{\begin{eqnarray}}
\newcommand{\ee}{\end{eqnarray}}
\newcommand{\nn}{\nonumber}

\newcommand{\Jcl}{\widetilde{J}}

\newcommand{\Nbar}{N_{0-}}

\newcommand{\omp}{\omega_{1+}}
\newcommand{\omm}{\omega_{3-}}

\newcommand{\Jop}{J_{0+}}
\newcommand{\Jom}{J_{0-}}
\newcommand{\Jonep}{J_{1+}}
\newcommand{\Jonem}{J_{1-}}
\newcommand{\Jtwop}{J_{2+}}
\newcommand{\Jtwom}{J_{2-}}
\newcommand{\Jthreep}{J_{3+}}
\newcommand{\Jthreem}{J_{3-}}
\newcommand{\Np}{N_{0+}}
\newcommand{\Nm}{N_{0-}}

\newcommand{\Jflat}{\mathcal{J}}

\newcommand{\F}{\mathcal{F}}

\newcommand{\vbar}{\bar{v}}

\def\al{\alpha}
\def\bt{\beta}

\def\btd{\dot{\beta}}
\def\ald{\dot{\alpha}}
\def\gmd{\dot{\gamma}}
\def\dld{\dot{\delta}}
\def\gm{\gamma}                
\def\dl{\delta}                
\def\eps{\epsilon}
\def\lm{\lambda}               
               
\def\om{\omega}               \def\Om{\Omega}

\def\si{\sigma}

\makeatletter
\def\mr@ignsp#1 {\ifx\:#1\@empty\else #1\expandafter\mr@ignsp\fi}%
\newcommand{\multiref}[1]{\begingroup
\xdef\mr@no@sparg{\expandafter\mr@ignsp#1 \: }%
\def\mr@comma{}%
\@for\mr@refs:=\mr@no@sparg\do{\mr@comma\def\mr@comma{,}\ref{\mr@refs}}%
\endgroup}
\makeatother

\newcommand{\hypref}[2]{\ifx\href\asklfhas #2\else\href{#1}{#2}\fi}

\newcommand{\secref}[1]{Sec.~\multiref{#1}}
\newcommand{\Appref}[1]{Appendix~\multiref{#1}}
\newcommand{\appref}[1]{App.~\multiref{#1}}

\newcommand{\figref}[1]{Fig.~\multiref{#1}}
\renewcommand{\eqref}[1]{(\multiref{#1})}

\newcommand{\half}{\frac{1}{2}}

\begin{document}

\thispagestyle{empty}
\begin{flushright}\footnotesize
\texttt{arxiv:0808.0282}\\
\texttt{UUITP-17/08} \vspace{0.8cm}
\end{flushright}

\renewcommand{\thefootnote}{\fnsymbol{footnote}}
\setcounter{footnote}{0}

\begin{center}
{\Large\textbf{\mathversion{bold}
Aspects of quantum integrability \\
for pure spinor superstring in $AdS_5\times S^5$
}\par}

\vspace{1.5cm}

\textrm{Valentina Giangreco M. Puletti} \vspace{8mm}

\textit{
Department of Physics and Astronomy,\\
Uppsala University,\\
SE-75108 Uppsala, Sweden}\\
\texttt{Valentina.Giangreco@fysast.uu.se} \vspace{3mm}


\par\vspace{1cm}

\textbf{Abstract} \vspace{5mm}

\begin{minipage}{13cm}
We consider the monodromy matrix for the pure spinor IIB superstring on $AdS_5\times S^5$ at leading order at strong coupling, in particular its variation under an infinitesimal and continuous deformation of the contour. Such variation is equivalent to the insertion of a local operator. Demanding the BRST-closure for such an operator rules out its existence, implying that the monodromy matrix remains contour-independent at the first order in perturbation theory. Furthermore we explicitly compute the field strength corresponding to the flat connections up to leading order and directly check that it is free from logarithmic divergences. The absence of anomaly in the coordinate transformation of the monodromy matrix and the UV-finiteness of the curvature tensor finally imply the integrability of the pure spinor superstring at the first order.
\end{minipage}

\end{center}

\vspace{0.5cm}

\newpage
\setcounter{page}{1}
\renewcommand{\thefootnote}{\arabic{footnote}}
\setcounter{footnote}{0}

\tableofcontents


\section{Introduction and Summary}
\label{sec:intro}

The AdS/CFT correspondence \cite{Maldacena:1997re, Gubser:1998bc, Witten:1998qj} realizes the holographic principle between type IIB superstring in $AdS_5\times S^5$ and $\mathcal N=4$ Super Yang-Mills theory. Providing a complete proof of the duality is a hard task due to the strong/weak coupling nature of the correspondence. In the planar limit integrability is playing a key role in this perspective. Both sides of the duality, gauge theory \cite{Minahan:2002ve, Beisert:2003tq, Beisert:2003yb} and string theory \cite{Mandal:2002fs, Bena:2003wd, Kazakov:2004qf}, manifest integrable structures and the assumption of an exact integrability has allowed to reach enormous progresses through the asymptotic S-matrix and the Bethe Ansatz machinery \cite{Beisert:2006ez, Eden:2006rx, Beisert:2006ib, Staudacher:2004tk, Beisert:2005tm, Beisert:2005fw, Arutyunov:2004vx, Arutyunov:2006iu}
\footnote{
For reviews in the vast subject of integrability we refer to \cite{Beisert:2004ry, Zarembo:2004hp, Plefka:2005bk, Minahan:2006sk} and references therein.}.

From a string theory point of view while the classical integrability of type IIB superstring in $AdS$ space was proved in \cite{Mandal:2002fs, Bena:2003wd} for the Metsaev-Tseytlin (MT) formulation \cite{Metsaev:1998it} and in \cite{Vallilo:2003nx} for the pure spinor (PS) version \cite{Berkovits:2000fe}, for quantum integrability there have been numerous evidences from various approaches however it is still substantially a conjecture. In this work we want to follow a more direct approach in order to check integrability at quantum level for the pure spinor IIB superstring. Along this line in \cite{Puletti:2007hq}, integrability has been directly checked in the near-flat-space limit \cite{Maldacena:2006rv}  at one-loop showing the factorization of the full three-particle S-matrix; in the PS formalism strong hints have been given in \cite{Mikhailov:2007mr} explicitly showing that the one-loop monodromy matrix is free from logarithmic divergences, and recently in \cite{Linch:2008nt} quantum conservation for the non-local charges has been showed for the gauged linear sigma model proposed by Berkovits and Vafa \cite{Berkovits:2007rj}.
%

Our goal, as mentioned above, is to explore the integrability of the type IIB superstring in $AdS_5\times S^5$ at the first order using the Berkovits formalism.
The main advantage in the PS approach is the covariant formulation which allows to quantize the string world-sheet action without spoiling the $D=10$ supersymmetry.
As in the Green-Schwarz (GS) formalism \cite{Green:1983wt} the target space supersymmetries are manifest and as in the GS Metsaev-Tseytlin action \cite{Metsaev:1998it}, the Berkovits action \cite{Berkovits:2000fe} is formulated in terms of the Maurer-Cartan forms. However the new ingredients are the bosonic ghosts $\lm_3\;,\lm_1$ (with their conjugate fields $\om_{1+}\;,\om_{3-}$) which are constrained to satisfy the so-called \textit{pure spinor constraints}
\[
\label{eq:ps-constraint}
\lm_1\gm^\mu\lm_1= \lm_3\gm^\mu\lm_3=0\;,
\]
with $\gm^\mu$ the $SO(9,1)$ gamma matrices\footnote{For more details we refer the reader to \appref{app:notation}. }, and the BRST-like operator Q
\[
Q= \int \textrm{Str}\big( \lm_1 \Jthreem  + \lm_3\Jonep\big)\;,
\]
which replaces the local fermionic $\kappa$-symmetry in the MT action.

In this work we show directly that the monodromy matrix remains independent of the contour at the first order in perturbation theory implying that the PS superstring in $AdS_5\times S^5$ is quantum integrable at one-loop and giving strong suggestions that it should be fully quantum integrable. This is done by considering an ansatz for the most general possible operator $\mathcal O$ which might give rise to an anomaly in the coordinate transformation of the monodromy matrix and proving that such operator $\mathcal O$ does not exist. The key point is that $\mathcal O$ must satisfy various conditions, in particular it must transform properly under the action of the BRST operator $Q$ \eqref{brst}. Eventually this requirement turns out to be the most strict one and we are able to prove that there are no operators satisfying this requirement. This implies the absence of anomaly.


In the second part of the work we explicitly compute $\F$ at leading order showing that indeed all the logarithmic divergences cancel without affecting the field strength.

\paragraph{Outline.} The work is organized as follows. In the next section (\secref{sec:preliminaries}) we review briefly the PS formulation for the type IIB superstring and the basic concepts of integrability. In \secref{sec:no-anomaly} we give a proof for the absence of operators $\mathcal O$ which are possible sources of anomalies and more details about this computation are in \appref{app:brst}. We compute the effective action and the OPE's for the currents in \secref{sec:effective-action}. In \secref{sec:field-strength} we explain how to compute the field strength $\F$ and we report two examples in order to show how the divergent terms in $\F$ cancel, other three cases are contained in \appref{app:comm-results}. In \secref{sec:conclusions} we summarize our results. The first appendix (\appref{app:notation}) contains all the details concerning notation and conventions adopted. In \appref{app:OPE-results} we list all the results for the OPE's necessary for the computation of $\F$. Finally some useful algebraic identities are contained in \appref{app:alg-identities}.
%


\section{Preliminaries}
\label{sec:preliminaries}

In the next section we summarize the PS formulation, for more details about the conventions and the notation used we refer the reader to the \Appref{app:notation}.

\subsection{Action and Equations of motion}
\label{sec:action}

The action for the Type IIB superstring in $AdS_5\times S^5$ with Ramond-Ramond (RR) flux in the PS formalism is \cite{Berkovits:2000fe, Berkovits:2000yr, Vallilo:2002mh}
\be
\label{initialaction}
S= \frac{R^2}{\pi} \int d^2z \;\textrm{Str}\big(
    \frac{1}{2} \Jtwop\Jtwom +\frac{3}{4}\Jonep\Jthreem + \frac{1}{4}\Jthreep\Jonem + \nn \\
    \omp \pbar \lm_3+ \omm\p\lm_1+ \Np\Jom + \Nm \Jop -\Nm\Np \big)\;,
\ee
where $R$ is the common radius for $S^5$ and $AdS_5$.
$\Np$ and $\Nm$ are the $SO(4,1)\times SO(5)$ components of the ghost Lorentz currents
\[
\label{def-ghost-current}
\Np =-\{\omp,\lm_3\}
\qquad
\Nm=-\{\omm,\lm_1\}
\]
and the right-invariant ''matter'' currents are
\[
\label{def-current}
J_+ =-\p g g^{-1}
\qquad
J_-=-\pbar g g^{-1},
\qquad
g\equiv h g
\]
where $g(x,\vartheta_L,\vartheta_R)$ parameterizes the super-coset ${PSU(2,2|4)\over SO(4,1)\times SO(5)}$ and $h$ the local $SO(4,1)\times SO(5)$ transformations.
The $\mathfrak{psu}(2,2|4)$ super Lie-algebra has a $\bold{Z}_4$ inner symmetry \cite{Berkovits:1999zq}, which decomposes it in
\[
\mathfrak{psu}(2,2|4)\equiv\mathfrak{g} = \mathfrak{g}_0 +\mathfrak{g}_1+\mathfrak{g}_2+\mathfrak{g}_3\;.
\]
$\mathfrak{g}_0 +\mathfrak{g}_2$ are the bosonic subalgebras, in particular $\mathfrak{g}_0$ is the $\bold{Z}_4$-invariant subalgebra for the gauge group $SO(4,1)\times SO(5)$, $\mathfrak{g}_2$ contains the remaining bosonic elements, while the fermionic subalgebras are $\mathfrak{g}_1+\mathfrak{g}_3$. Consequently the matter currents $J_{\pm}$ decompose as
\[
J_{\pm}=J_{\pm}^A T_A=
J_{0\pm}^{[\mu\nu]}t_{[\mu\nu]}^0+
J_{1\pm}^{\ald}t_{\ald}^1+
J_{2\pm}^\mu t_\mu^2 +
J_{3\pm}^\al t_\al^3\;,
\]
where $t_A$ are the $\mathfrak{psu}(2,2|4)$ generators.

The action \eqref{initialaction} is classically invariant under BRST transformation generated by Q
\[
\label{Q-brst}
Q= \int \textrm{Str}\big( \lm_1 \Jthreem  + \lm_3\Jonep\big)\;
\]
and \eqref{initialaction} has been proved to be BRST and conformal invariant also at quantum level
in \cite{Berkovits:2004xu}. Conformal invariance was also checked explicitly in \cite{Vallilo:2002mh} at one-loop.


\paragraph{Equations of motion.}
Under small variations $\xi\in \mathfrak{g}_i$ with $i=1,2,3$ of the fields $g$ the currents satisfy
\[
\label{first-var-j}
\dl_\xi J_+ = -\p \xi -[J_+,\xi]
\qquad
\dl_\xi J_- =-\pbar \xi -[J_-,\xi]\;.
\]
Plugging \eqref{first-var-j} into the action \eqref{initialaction} and using the Maurer-Cartan identities
$
\p J_- -\pbar J_+ +[J_+,J_-]=0\;,
$
one obtains the following equations of motion for the matter currents
\be
\label{eom-matter}
&& D_+ \Jtwom +[\Jthreep,\Jthreem] -[\Np,\Jtwom]+[\Jtwop,\Nm]=0 \nn \\
&& D_- \Jtwop + [\Jonem,\Jonep] - [\Np,\Jtwom] + [\Jtwop,\Nm]=0 \nn \\
&& D_+\Jthreem + [\Jthreem,\Np] + [\Jthreep,\Nm]=0 \nn \\
&& D_- \Jthreep +[\Jtwom,J_1]+[\Jonem,\Jtwop]+[\Jthreem,\Np]-[\Nm,\Jthreep]=0 \nn \\
&& D_- \Jonep + [\Jonem,\Np]-[\Nm,\Jonep]=0 \nn \\
&& D_+ \Jonem + [\Jtwop,\Jthreem]+ [\Jthreep,\Jtwom]+[\Jonem,\Np]-[\Nm,\Jonep]=0 \;,
\ee
where the covariant derivatives are $D_+=\p+[\Jop,\;]$, $D_-=\pbar+[\Jom,\;]$.

Analogously we can derive the equations of motion for the ghost fields $\lm$ and $\om$, namely
\be
\label{eom-ps}
&&D_-\lm_3-[\Nm,\lm_3]=0
\qquad
D_+\lm_1-[\Np,\lm_1]=0
\\ \nn
&& D_-\om_{1+}-[\Nm,\om_{1+}]=0
\qquad
D_+\om_{3-}-[\Np,\om_{3-}]=0 \;,
\ee
together with the pure spinor constraints \eqref{eq:ps-constraint}, which can be rewritten as
\[
[\lm_3,\Np]=0
\qquad
[\lm_1,\Nm]=0\;.
\]
From \eqref{eom-ps} it follows the equations of motion for the ghost currents, i.e.
\be
\label{eom-ghost}
&& D_+ \Nm -[\Np,\Nm]=0 \nn \\
&& D_- \Np - [\Nm,\Np]=0\;.
\ee


\paragraph{BRST transformation.}

The coset representative $g(x,\vartheta_L,\vartheta_R)$ transforms under BRST action as \cite{Berkovits:2004jw}
\[
\eps Q(g)=g (\eps \lm_1+\eps \lm_3)
\]
which implies for the currents the following expressions
\be
\label{brst-transf}
&& \eps Q (J_{i+}) =
     \dl_{i+3,0}\p (\eps \lm_1)+[J_{i+3;\,+},\eps\lm_1]+
     \dl_{i+1,0}\p (\eps \lm_3)+[J_{i+1;\,+},\eps\lm_3]\nn \\[1mm]
&& \eps Q (J_{i-}) =
     \dl_{i+3,0}\pbar (\eps \lm_1)+[J_{i+3;\,-},\eps\lm_1]+
     \dl_{i+1,0}\pbar (\eps \lm_3)+[J_{i+1;\,-},\eps\lm_3]
\nn \\[1mm]
&& \eps Q (N_+)=[\Jonep,\lm_3]
\qquad
\eps Q (N_-)=[\Jthreem,\lm_1]\;,
\ee
where $i=1,2,3$ labels the corresponding subalgebras, i.e. $J_i\equiv J_{|\mathfrak{g}_i}$.


\subsection{Flat connections and monodromy matrix}
\label{sec:classical-int}

Models which have infinitely many conserved charges are integrable. In the PS formalism classical integrability was studied in \cite{Vallilo:2003nx, Berkovits:2004jw, Bianchi:2006im, Kluson:2006wq}. In this perspective the central point is the construction of flat connections, namely a linear combination of ghost and matter currents which satisfies the zero-curvature equation and which is parameterized by a complex parameter $z$ (\textit{spectral parameter}). For the PS superstring such flat connections (\textit{Lax pair}) were constructed in \cite{Vallilo:2003nx}
\be
\label{def-flat-connection}
\mathcal{J}_+(z) &=& \Jop +z \Jtwop + z^{1/2} \Jthreep + z^{3/2} \Jonep + (z^2 -1)\Np \cr
&&\cr
\mathcal{J}_-(z) &=& \Jom + z^{-1}\Jtwom + z^{-3/2}\Jthreem + z^{-1/2}\Jonem+ (z^{-2}-1)\Nm\;.
\ee
Indeed using the equations of motion \eqref{eom-matter}, \eqref{eom-ghost} one can see that the corresponding field strength
\[
\label{def-f}
\F_{+-}(z)\equiv \p \mathcal{J}_-(z) -\pbar\mathcal{J}_+(z)+[\mathcal{J}_+(z), \mathcal{J}_-(z)]
\]
vanishes, namely
\[
\label{zero-curv-eq}
\F_{+-}(z)=0\;.
\]
The zero-curvature equation \eqref{zero-curv-eq} (\textit{Lax equation}) encodes all informations about the equations of motion and the Maurer-Cartan identities.

From the connections $\mathcal J_{\pm}$ one can construct a Wilson-like operator (\textit{ monodromy matrix}) as
\be
\label{monodromy}
\Omega(z)=\mathrm{P}\exp \oint_{\mathcal{C}} \mathcal{J}(z)\;,
\ee
where $\mathrm{P}$ indicates the path-ordering prescription. Notice that $\Jflat$ takes value in the $\mathfrak{psu}(2,2|4)$ super algebra, while $\Om$ is a supergroup-valued matrix.

Furthermore the action \eqref{initialaction} and the flat connections \eqref{def-flat-connection} (consequently also the monodromy matrix) are symmetric under the following parity transformation \cite{Berkovits:2004xu}
\[
\label{symm-f}
z \leftrightarrow z^{-1}\qquad \text{holomorphic}\leftrightarrow \text{anti-holomorphic}\qquad \mathfrak{g}_1\leftrightarrow \mathfrak{g}_3\;.
\]

Since $\mathcal J$ satisfies the zero-curvature equation \eqref{zero-curv-eq} at classical level, then the monodromy matrix $\Omega$ \eqref{monodromy} is classically independent of the shape of the path. In general the variation of a Wilson loop operator caused by the infinitesimal deformation of the contour is given by \cite{Polyakov:1980ca}
\be
\label{var-coord-mm}
\frac{\delta}{\delta x^a(s)}\Omega= \mathrm{P} \big(\mathcal{F}_{ab}\dot x^b(s) \exp\oint_{\mathcal{C}} \mathcal{J}(s)\big)\;,
\ee
where $\mathcal{F}_{ab}$
\footnote{
Here for brevity we adopt a covariant notation for the coordinates, i.e. $a, b$ label the world-sheet coordinates.
}
is the field strength corresponding to $\mathcal{J}_a$ and $s$ parameterizes the contour $\mathcal C$. This indeed is another way of saying that if the zero-curvature equation holds \eqref{zero-curv-eq}, then we can continuously deform the path in $\Omega$ without producing any effect.

Consequently $\Omega$ can be used to generate an infinite set of conserved charges: the independence of the contour for the monodromy matrix is equivalent to the conservation of the charges.
For example the charges can be obtained as a Taylor expansion of the super-trace of $\Omega(z)$ and in particular the expansion around different values of the spectral parameter ($z=0,\infty$ or $z=1$) gives the conserved charges, local or non-local respectively.


\section{Absence of anomaly}
\label{sec:no-anomaly}

We now want to move to the quantum theory. Quantum integrability for the PS superstring was studied in \cite{Berkovits:2004xu, Adam:2007ws, Mikhailov:2007eg, Mikhailov:2007mr}. A theory which is classically integrable not necessarily will be quantum integrable \cite{Abdalla:1982yd, Luscher:1977uq}. A famous example in literature is the $\mathbb{C}P^n$-model \cite{Abdalla:1980jt}.
At quantum level the currents $\mathcal J$ become composite operators, and both the monodromy matrix $\Omega$ \eqref{monodromy} and the field strength $\mathcal F$ \eqref{def-f} contain their product. Typically the product of operators is not well-defined. Thus in general the current short-distance behavior produces divergences, which might spoil the classical conservation laws, giving rise to anomalies  and making the quantum theory not integrable.

Motivated by the results in \cite{Mikhailov:2007mr}, where as already mentioned, the monodromy matrix $\Omega$ \eqref{monodromy} was shown to be free from the one-loop divergences, here we want to investigate the deformation of $\Om$ under infinitesimal variations of the contour $\mathcal C$.
Such variations correspond to an insertion of a local operator and if $\dl\Om$ is not zero, namely if there is an anomaly, this implies that there exists a non-vanishing operator $\mathcal O$ sitting in this infinitesimal deformed path.

This operator $\mathcal O$ will be local since as explained above, we need to worry about the short-distance behavior, and by dimensional analysis it is expected to have conformal dimension $(1,1)$ \cite{Abdalla:1980jt, Luscher:1977uq}. Moreover since the Wilson loop is BRST-invariant at classical level \cite{Berkovits:2004jw} and at all orders in perturbation theory \cite{Berkovits:2004xu}, then also $\mathcal O$ must obey to the equation
\be
\label{brst}
Q \cdot\mathcal{O}^{(1,1)}= [\mathcal{O}^{(1,1)},z^{1/2}\lambda_3+z^{-1/2}\lambda_1]\;.
\ee
Moreover from \eqref{brst} it follows that it should have ghost number zero.

If such an operator $\mathcal O^{(1,1)}$%
\footnote{I thank A. Mikhailov since the results of this section benefited from his numerous and valuable suggestions.}
exists then it is constructed from the ghost and matter currents, in particular it cannot contain $J_{0\pm}$ since it has to be gauge invariant, and it cannot contain $J_{3+}$ and $J_{1-}$ because their BRST transformations \eqref{brst-transf} produce also ghost derivatives while the equation \eqref{brst} needs to be satisfied exactly and not up to derivatives. Following the same technique in \cite{Berkovits:2004xu} we can  write $\mathcal O^{(1,1)}$ as a linear combination of the form
\be
\label{O-def}
\mathcal{O}^{(1,1)}\,(z) &=&
 A^{2+,2-}(z)[J_{2+},J_{2-}]
+ A^{1+,3-}(z)[J_{1+},J_{3-}]
+ A^{2+,3-}(z)[J_{2+},J_{3-}] \nn \\[1mm]
&+&
 A^{1+,2-}(z)[J_{1+},J_{2-}]
+ A^{0+,2-}(z)[\Np,\Jtwom]
+ A^{0-,2+}(z)[J_{2+},\Nm]\nn \\[1mm]
&+&
A^{1+,0-}(z)[J_{1+},\Nm]
+ A^{0+,3-}(z)[\Np,\Jthreem]
+ A^{0+,0+}(z)[\Np,\Nm] \;. \nn \\[1mm]
\ee
The coefficients $A$ are arbitrary functions of the spectral parameter $z$ and at the leading order they are of order $h$. The currents in \eqref{O-def} satisfy the Maurer-Cartan identities and the classical equations of motion. That is why we do not need to include also terms with the derivatives of the currents in \eqref{O-def} since they will be related to the commutators by the equations of motion%
\footnote{
There is a further constraint in the coefficients of the expression \eqref{O-def} coming from the non perturbative symmetry \eqref{symm-f}. Since the commutators are antisymmetric with respect to the parity transformation \eqref{symm-f}, then we should also require that
\be
\label{symmetry-A}
A^{2+,2-}(z)= A^{2+,2-}(z^{-1})
\cr
A^{2+,3-}(z)= A^{1+,2-}(z^{-1})
\ee
and so on for the other functions. However we do not need to use the equations \eqref{symmetry-A} in order to solve the relations \eqref{system-A}.
}.
Imposing the equation \eqref{brst} to the operator $\mathcal O^{(1,1)}$ \eqref{O-def} and using the BRST transformations \eqref{brst-transf},
it is straightforward to obtain the following system of linear equations for the arbitrary functions $A$
\be
\label{system-A}
A^{0+,3-}= A^{1+,0-} = A^{0+,0-}
\qquad
A^{1+,0-}= A^{2+,3-}
\nn \\[1mm]
A^{1+,3-}=A^{1+,2-}=A^{2+,3-}
\qquad
A^{0+,3-}=A^{1+,2-}
\nn \\[1mm]
A^{2+,2-}=A^{2+,3-}=A^{1+,2-}
\qquad
A^{1+,2-}=A^{0+,2-}
\nn \\[1mm]
A^{2+,0-}=A^{2+,3-}
\qquad
A^{2+,3-}=A^{1+,2-}=0
\nn \\[1mm]
A^{2+,0-}=0
\qquad
A^{0+,2-}=0\;.
\ee

In \appref{app:brst} more details about the computation of the system \eqref{system-A} are given. All the coefficients in \eqref{system-A} are related and the last conditions exclude any possible non-trivial solution for the system. This means that no operator $\mathcal O$, satisfying all the properties listed above, exists, ruling out the anomaly in the quantum monodromy matrix under path deformations%
\footnote{
We also tried combinations of operators which include finite terms such as those described in \secref{sec:OPE}. However whenever we demand that equation \eqref{brst} is satisfied, this excludes any possibility to find a solution for the system of type \eqref{system-A}.
}.

The validity of the equation \eqref{brst} is indeed much stronger, since in \cite{Berkovits:2004xu} Berkovits showed that the non-local charges are BRST invariant at all orders and he constructed the local counter-terms in order to take into account quantum effects for the BRST operator. Thus we can still write $\mathcal O$ as in \eqref{O-def} at any $n$-loop order in a quantum theory with the unknown functions of order $h^n$.

In performing this analysis we basically mod out in the expression of $\mathcal O$ \eqref{O-def} the redundancy coming from the Maurer-Cartan equations and from the equations of motion. Hence in our space of possible operators satisfying all the conditions discussed above (local, dimension-two, ghost number zero and BRST-closed), we are considering the restricted set of operators which are not zero on-shell. However we keep in mind that we could consider for example $\mathcal O =\mathcal F$, where the prototype of such operator $\F$ is the field strength. Trivially one can check that $\F$ satisfies all the listed constraints, but it vanishes classically \eqref{zero-curv-eq}.

The situation is different in the quantum $\mathbb{C}P^n$ model \cite{Abdalla:1980jt}.
In that case there is no analogue of the constraint \eqref{brst}, and in the absence
of such a constraint there is in fact
an operator ${\cal O}$ with the right conformal dimension $(1,1)$, giving rise to an anomaly \cite{Abdalla:1980jt}.
In our model, as we have explained, \eqref{brst} implies that  the anomaly vanishes.

It has been explained in \cite{Mikhailov:2007mr} that the independence of the contour for the monodromy matrix
implies the cancellation of the logarithmic divergences. Therefore our argument also implies the finiteness of the
transfer matrix to all orders of $\alpha'$.

In order to make our statement stronger, in the second part of the paper we explicitly compute the variation of the monodromy matrix at the leading order. According to our argument, it should be zero.
Because of the technical difficulties, we have not completely demonstrated the cancellation,
but we do demonstrate the cancellation of the log divergences in the field strength. (What we have not
explicitly demonstrated is that the finite terms also cancel.)



\section{Effective action and OPE's}
\label{sec:effective-action}

In this part of the work the goal is to investigate the equation \eqref{zero-curv-eq} at first order in perturbation theory by computing explicitly the current short-distance behavior, i.e. their OPE's, and the field strength $\F$. This is done using the background field method \cite{Berkovits:1999zq}. The expansion parameter naturally is ${1 \over R}$ with $R\rightarrow \infty$ and the analysis is valid up to ${1\over R^2}$.


\subsection{Effective action and vertices}

In order to compute the OPE's for the currents contained in \eqref{zero-curv-eq} one needs to know the interaction vertices, namely the effective action for the quantum fluctuations.
For practical reasons we treat separately the terms containing only matter currents from the interactions containing matter and ghost currents.

\subsubsection{Matter vertices}
\label{sec:matter-vertex}

The super-group-valued map $g$ is expanded in the quantum fluctuations $X \in \mathfrak{g/g_0}$ around the classical point $\tilde g$, i.e. $g=\exp{({1\over R} X)}\;\tilde{g}$. Consequently the matter currents $J =-d g g^{-1}$ become
\be
\label{Jmatter-exp}
J_{i\pm} &=&
      \Jcl_{i\pm}+{1\over R}J_{i\pm}^{(1)}+{1\over R^2}J_{i\pm}^{(2)}+...=\\ \nn
     &=& \Jcl_{i\pm}
     -{1\over R}\big([\Jcl_{\pm},X]_i+\partial_{\pm} X_i\big)
     +{1\over 2 R^2}\big([[\Jcl_{\pm},X],X]_i+[\partial_{\pm} X,X]_i\big)+...\;,
\ee
where $i= 1,2,3$ and $\Jcl$ denotes the classical current $\Jcl =-d \tilde g \tilde g^{-1}$.
Also the gauge fields can have quantum fluctuations and the corresponding expansion is
\be
\label{Jzero-exp}
J_{0\pm} &=&
         \Jcl_{0\pm} +{1\over R}J_{0\pm}^{(1)}+{1\over R^2}J_{0\pm}^{(2)}+...=\\ \nn
         &=&
     \Jcl_{0\pm}
    -{1\over R} [\Jcl_{\pm},X]_0
    +{1\over 2 R^2} \big([[\Jcl_{\pm},X],X]_0+[\partial_{\pm} X,X]_0\big)+...
\;.
\ee
Inserting the expansions \eqref{Jmatter-exp} and \eqref{Jzero-exp} in the action one obtains terms of zeroth order in the $X$ fields, which is the classical action, terms of first order in X, which vanish by classical equations of motion, and finally quadratic terms in X.
We need to take into account the interactions which are quadratic in the $X$ fields, namely the interactions of order ${1\over R^2}$. Since the effective action is invariant under gauge transformations the gauge can be further fixed such that $[\Jom,X_i]=[\Jop,X_i]=0$~\cite{Berkovits:1999zq}.

Plugging the expansions \eqref{Jmatter-exp} and \eqref{Jzero-exp} for the currents in the matter action
\[
\label{matter-action-cl}
S_M=\frac{R^2}{\pi} \int d^2z \textrm{Str}\big( \frac{1}{2} \Jtwop\Jtwom +\frac{3}{4}\Jonep\Jthreem +\frac{1}{4}\Jthreep\Jonem\big)
\]
one obtains
\[
S_M=S_{M; 0}+ S_{M;\bt}+ S_{M;2}
\]
where $S_{M;0}$ is the classical matter action, $S_{M;\bt}$ is the effective action for the matter contribution used for computing the one-loop $\beta$-function in~\cite{Berkovits:1999zq} and in \cite{Vallilo:2002mh}, while $S_{M;2}$ contains terms which in principle can contribute now.

\begin{figure}
\begin{center}
\includegraphics[scale=1]{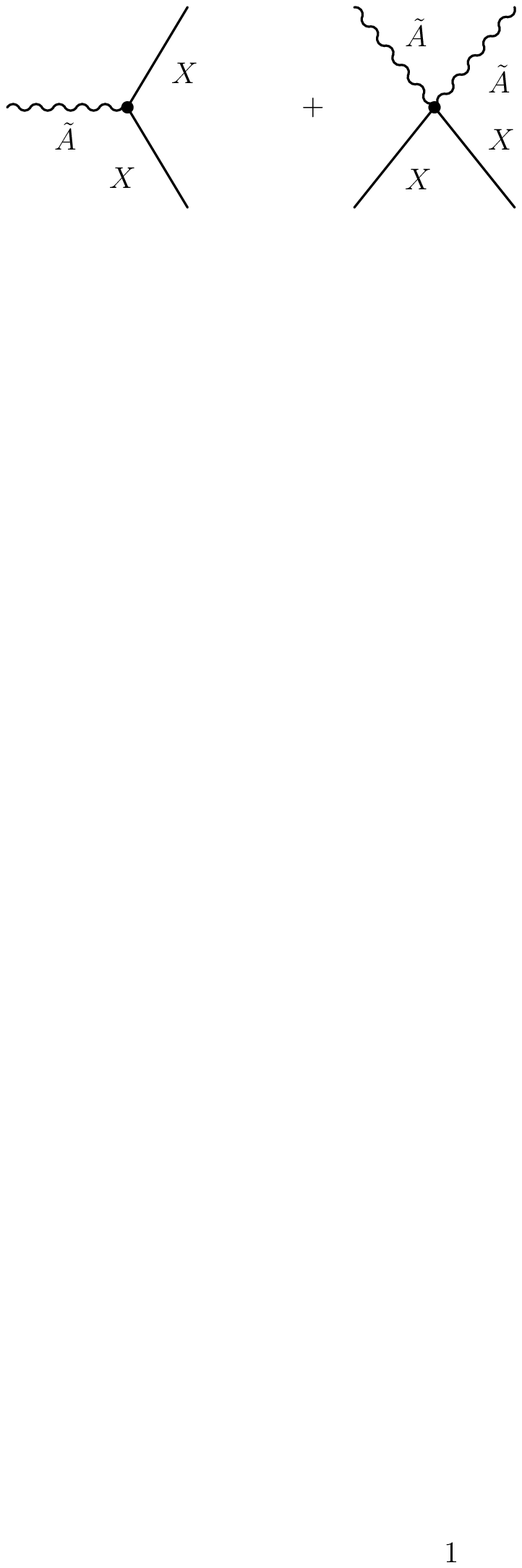}
  \caption{\textbf{Tree-level matter vertices.} The label $\tilde A$ indicates the various classical currents, i.e. ghost and matter currents. The label $X$ indicates the quantum matter fluctuations.}
\label{fig:tree-vertex}
\end{center}
\end{figure}

Explicitly:
\be
    S_{M;\bt}= \frac{1}{\pi} \int d^2z \;\textrm{Str}
        \big(
             \pbar X_3 \p X_1 + \half \pbar X_2 \p X_2  \\ \nn
            - [\p X_2, X_1] \Jonem - [\pbar X_2, X_3] \Jthreep
            - \frac{1}{2}[\p X_1, X_1] \Jtwom-\frac{1}{2} [\pbar X_3, X_3] \Jtwop \\ \nn
            + \frac{3}{4}[[\Jthreem, X_1], X_3] \Jonep + \frac{1}{2}[[\Jthreem, X_2], X_2] \Jonep +\frac{1}{4}[[\Jthreem, X_3],X_1] \Jonep  \\ \nn
            +\frac{1}{2}[[\Jtwom, X_2],X_2] \Jtwop + \frac{1}{4}[[\Jtwom, X_3], X_1] \Jtwop -\frac{1}{4} [[\Jtwom, X_1], X_3] \Jtwop \\ \nn
            -\frac{1}{4}[[\Jonem, X_1], X_3] \Jthreep -\frac{1}{2}[[\Jonem, X_2], X_2] \Jthreep+  \frac{1}{4}[[\Jonem, X_3], X_1]\Jthreep
        \big)
\ee
\be
\label{double-ins-mm}
    S_{M;2}= \frac{1}{\pi} \int d^2z \;\textrm{Str}
        \big(
         \frac{1}{2} [[\Jonem,X_3],X_3]\Jonep+ \half [[\Jthreem,X_1],X_1]\Jthreep \\ \nn
        + \frac{5}{8}[[\Jtwom,X_2],X_3]\Jonep+ \frac{3}{8}[[\Jtwom,X_3],X_2]\Jonep
        +\frac{3}{8}[[\Jthreem,X_2],X_1]\Jtwop \\ \nn
        +\frac{5}{8}[[\Jthreem,X_1],X_2]\Jtwop
      - \frac{3}{8}[[\Jonem,X_2],X_3]\Jtwop +\frac{3}{8}[[\Jonem,X_3],X_2]\Jtwop\\ \nn
      -\frac{3}{8}[[\Jtwom,X_1],X_2]\Jthreep+\frac{3}{8}[[\Jtwom,X_2],X_1]\Jthreep
      \big)
\ee
Notice that all the currents which appear in the effective action are the classical ones, the symbol $\;\widetilde{}\;$ is omitted. Thus we have two types of vertices as it is shown in \figref{fig:tree-vertex}.
%


\subsubsection{Ghost-Matter vertices}
\label{sec:ghost-vertex}

The background field method is applied also to the ghosts~\cite{Vallilo:2002mh, Bedoya:2006ic, Chandia:2003hn}
\be
\omp\rightarrow \tilde{\om}_{1+} + {1\over R} \om_1
\qquad
\lm_3\rightarrow \tilde\lm_3 +{1\over R} \lm_3 \\
\omm\rightarrow \tilde{\om}_{3-}+{1\over R} \omm
\qquad
\lm_1\rightarrow \tilde{\lm}_1+{1\over R}\lm_1\;,
\ee
where $\tilde\lm_3\,,\tilde\lm_1\,,\tilde\om_{3-}\,,\tilde\om_{1+}$ are the classical fields. This leads to the following expression for the ghost Lorentz currents
\be
\label{N-exp}
N_{0+}=\tilde{N}_{0+}+{1\over R} N^{(1)}_{0+}+ {1\over R^2} N_{0+}^{(2)}+...  \nn\\
\Nbar=\widetilde{N}_{0-} +{1\over R} N_{0-}^{(1)}+{1\over R^2} N_{0-}^{(2)}+...\;,
\ee
with
\be
&& N^{(1)}_{0+}=-\{\omp,\tilde{\lm}_3\}-\{\tilde{\om}_{1+},\lm_3\}
\qquad
N_{0-}^{(1)}=-\{\omm,\tilde{\lm}_1\}-\{\tilde{\om}_{3-},\lm_1\}\nn \\
&& N_{0-}^{(2)}=-\{ \om_{3-},\lm_1\}
\qquad
N_{0+}^{(2)}=-\{\omp,\lm_3\} \;.
\ee

\begin{figure}
\begin{center}
\includegraphics[scale=1]{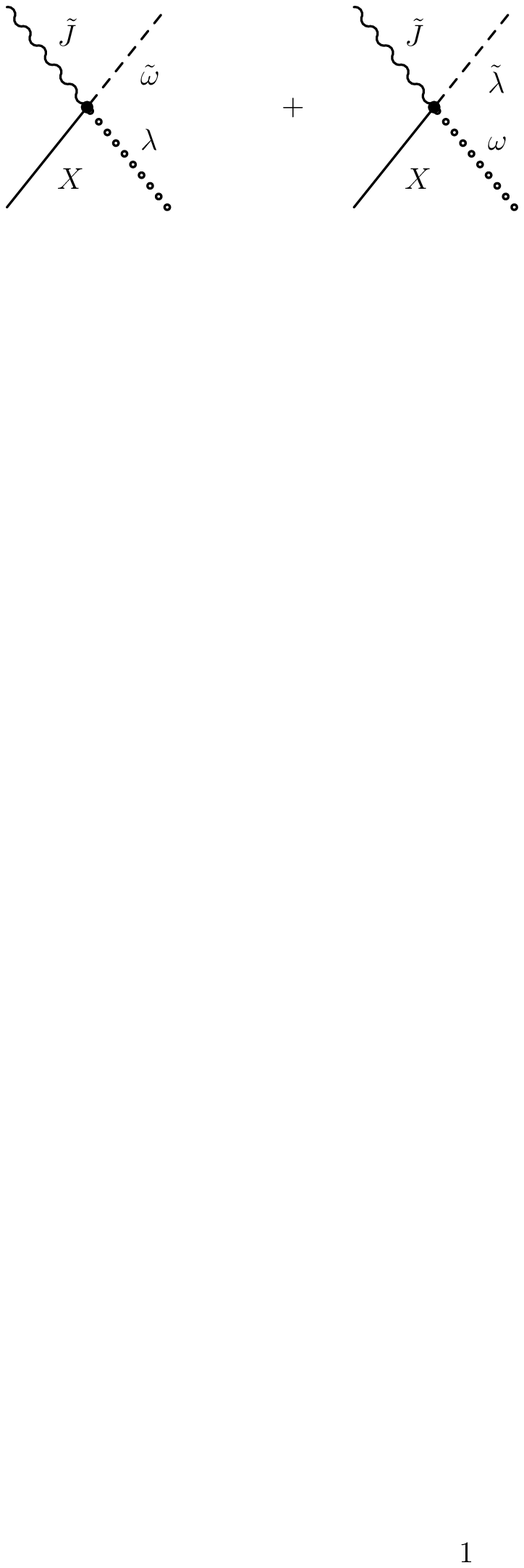}
\end{center}
  \caption{\textbf{Tree-level ghost-matter vertices.} The label $\tilde J$ indicates the classical matter current. The label $X$ indicates the quantum matter fluctuations. $\tilde\lm$ and $\tilde\om$ represent the classical ghosts, while $\lm\;,\om$ the quantum fluctuations for the ghost fields.}
\label{fig:tree-vertex-gm}
\end{figure}

The ghost-matter interactions are contained in
\[
\label{action-ghost-matter-cl}
S_{GM}= \frac{R^2}{\pi} \int d^2z \; \textrm{Str}\big( N_{0+}\Jom+\Nm J_{0+}\big)
\]
and using the expansion \eqref{N-exp}, \eqref{Jzero-exp} one obtains%
\[
S_{GM}= S_{GM;0}+ S_{GM;\bt}+ S_{GM;2}+S_{GM;3}\;.
\]
$S_{GM;0}$ is the classical ghost-matter action, $S_{GM;\bt}$ contributes to the one-loop $\bt$-function~\cite{Vallilo:2002mh}
\be
\label{three-leg-gm}
S_{GM;\bt}= \frac{1}{2\pi} \int d^2 z\; \textrm{Str}\big(
                        N_{0+} [\pbar X_1, X_3] +  N_{0+} [\pbar X_2, X_2] + N_{0+} [\pbar X_3, X_1]\\ \nn
                        + \Nm [\p X_1, X_3] + \Nm [\p X_2, X_2] + \Nm [\p X_3, X_1] \big)\;,
\ee
and further contributions are contained in
\be
\label{double-ins-mg}
S_{GM;2}= \frac{1}{2\pi} \int d^2z\;\textrm{Str}
           \big(
           N_{0+} [[\Jonem,X_1],X_2]+  N_{0+} [[\Jonem,X_2],X_1] \\ \nn
           + N_{0+} [[\Jtwom,X_1],X_1] + N_{0+} [[\Jthreem,X_3],X_2] +N_{0+} [[\Jthreem,X_2],X_3] \\ \nn
           + N_{0+} [[\Jtwom,X_3],X_3] + \Nm [[J_{1+},X_1],X_2]+  \Nm [[J_{1+},X_2],X_1] \\ \nn
           + \Nm [[J_{2+},X_1],X_1] +\Nm [[J_{3+},X_3],X_2] \\ \nn
           + \Nm [[J_{3+},X_2],X_3] + \Nm [[J_{2+},X_3],X_3]
           \big) \;,
\ee
\be
\label{4-leg-vertex-matter-ghost}
S_{GM;3}= \frac{1}{\pi} \int d^2z\; \textrm{Str} \big(
          - N^{(1)}_{0+}([\Jonem,X_3]+[\Jthreem,X_1]+[\Jtwom,X_2])\\ \nn
          - N_{0-}^{(1)}([J_{1+},X_3]+[J_{3+},X_1]+[J_{2+},X_2])
          \big)\;,
\ee
In particular $S_{GM;3}$ provides the four-leg-vertex between the ghosts and the matter fields (\figref{fig:tree-vertex-gm}), which will be responsible for the mixed OPE between $J$ and $N$.


\subsubsection{Ghost-Ghost vertices}
\label{sec:ghost-ghost-vertex}

\begin{figure}
\begin{center}
\includegraphics[scale=1]{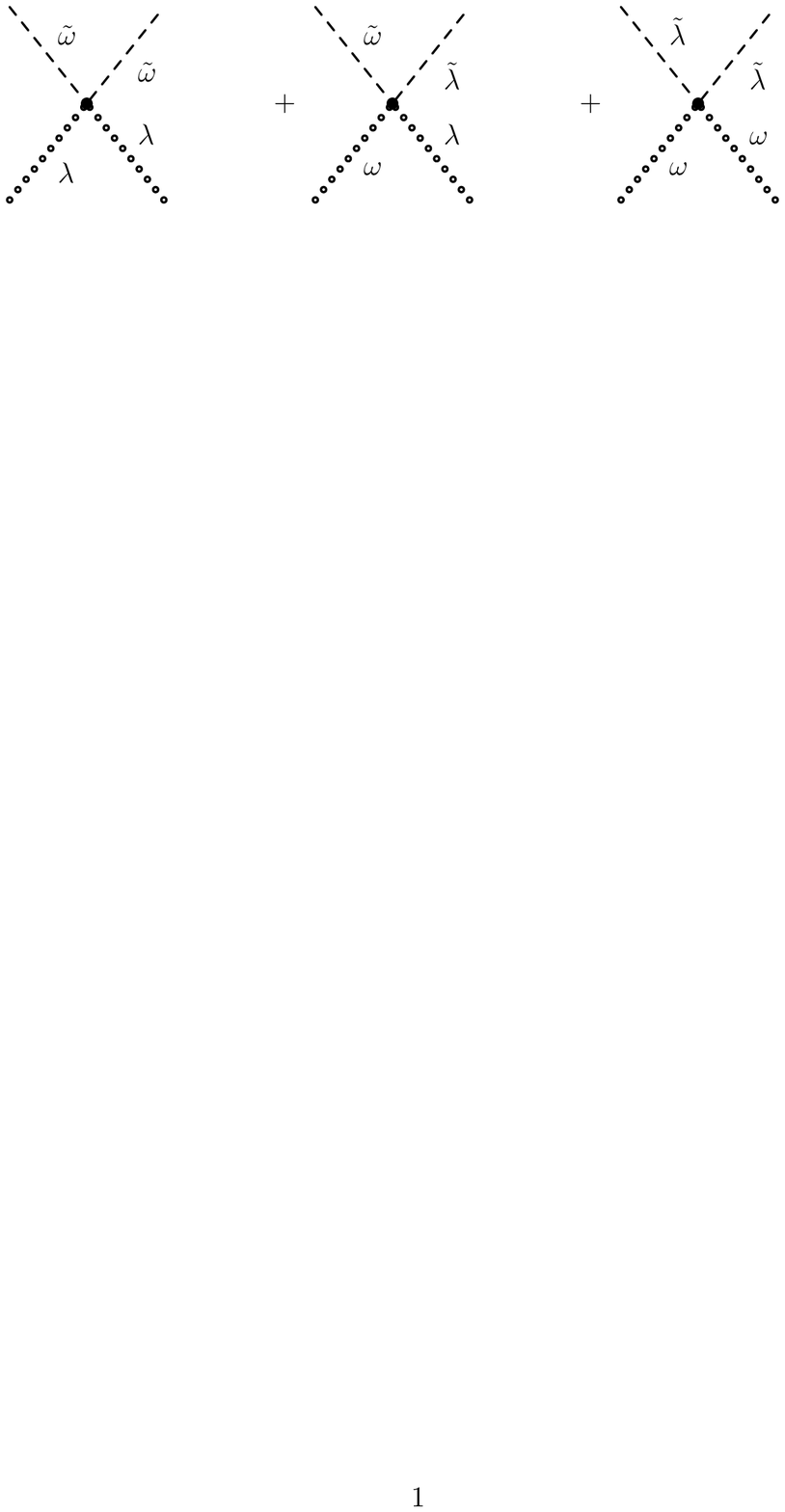}
\end{center}
  \caption{\textbf{Tree-level ghost vertices.} The $\;\widetilde{}\;$ labels the classical ghost fields. In each diagram all the two types of ghost appear, namely $\lm_1,\lm_3,\om_{1+},\om_{3-}$.}
\label{fig:tree-vertex-gg}
\end{figure}

The last contribution to the action is
\[
\label{action-ghost-ghost-cl}
S_{G}= -\frac{R^2}{\pi} \int d^2z\; \textrm{Str} \big( N_{0+}N_{0-}\big)\;.
\]
Using the expansion \eqref{N-exp}, it becomes
\be
S_{G}= S_{G;\,0}+ S_{G;2} \;,
\ee
with again $S_{G;\,0}$ the classical contribution and
\[
\label{eff-action-gg}
S_{G;2}= -\frac{1}{\pi} \int d^2z\; \textrm{Str} \big(
           N^{(1)}_{0+} N^{(1)}_{0-}\big)\;.
\]
$S_{G;2}$ is responsible for the interaction between the two types of ghost currents (\figref{fig:tree-vertex-gg}), so we will have also a non-zero OPE between $N_{0+}$ and $N_{0-}$%
\footnote{
In principle the effective one-loop action can have terms such as
\be
S_{GM;4}= \frac{1}{\pi} \int d^2z\; \textrm{Str} \big(
          N^{(2)}_{0+}\tilde J_{0-} + N_{0-}^{(2)}\tilde J_{0+}  \big)\;
\ee
or
\be
S_{G;4}= -\frac{1}{\pi} \int d^2z\; \textrm{Str} \big(
          N^{(2)}_{0+}\tilde N_{0-} + N_{0-}^{(2)}\tilde N_{0+}  \big)\;,
\ee
which could correct the propagators for the ghost fields. However, since at this order such corrections are not required, we do not enter in the details for the ghost propagators.}
.


\subsection{OPE's: general structure}
\label{sec:OPE}

The vertices obtained in \secref{sec:matter-vertex}, \secref{sec:ghost-vertex} and \secref{sec:ghost-ghost-vertex} correct the free propagators $A^{-1}$ using
\be
\label{prop-correction}
(A + V_1 + V_2)^{-1}&=&\\ \nn
&=& A^{-1}-(A^{-1}V_1 A^{-1})+(A^{-1}V_1 A^{-1}V_1 A^{-1}) -(A^{-1}V_2 A^{-1})+...\;,
\ee
where $A=\frac{1}{2\pi}(-\p\pbar)C_{AB}$, and $C_{AB}$ is $C_{\mu\nu}$ for the bosons and $C_{\al\ald}, C_{\btd\bt}$ for the fermions. In \eqref{prop-correction} we have distinguished the vertices with respect to the dimension of the operators. As one can see from the effective action and from the figures \figref{fig:tree-vertex}, \figref{fig:tree-vertex-gm} and \figref{fig:tree-vertex-gg}, the interaction terms that can be inserted, are essentially of two types:
\begin{enumerate}
\item $V_1$ which contains one classical current and a derivative acting on the propagators, i.e. $J\cdot\overrightarrow{\partial}+\overleftarrow{\partial}\cdot J$, (three-leg diagrams);
\item $V_2$ which contains two classical currents and basically is a multiplicative operator, i.e. $J\;J$, (four-leg diagrams).
\end{enumerate}
Since we are interested in dimension-two operators, we will consider up to vertices with two classical currents ($J\,J$) and with one derivative of the currents ($\partial\,J$).
Notice that for this reason, vertices of the first type $V_1$ can be Taylor-expanded.

There are four different types of dimension-two operators which are produced in the OPE's \eqref{opej}, \eqref{ope-nj} and \eqref{ope-nn}:
\begin{itemize}
\item $\partial_{\pm} J_{i\mp}\;, \partial_{\pm} J_{i\pm}$
\item $[J_{i\pm}, N_{0\mp}]\;, [J_{i\pm}, N_{0\pm}]$
\item $[J_{i\pm}, J_{j\mp}]\;,[J_{i\pm}, J_{j\pm}]$
\item $[N_{0\pm}, N_{0\mp}]\;, [N_{0\pm},N_{0\pm}]$,
\end{itemize}
where $i=1,2,3$. There is no $J_0$ since it is gauged away in the expansions \eqref{Jmatter-exp}, \eqref{Jzero-exp}, and in any case the result of the computation must be gauge-invariant.

Due to the Lorentz invariance these operators come with different space-time behaviors:
\begin{itemize}
\item Operators with one holomorphic and one anti-holomorphic components, e.g. $J_{i-} J_{j+}\;, N_{0+} N_{0-}\;\pbar J_{i+}\;, \p J_{i-}$, have logarithmic divergences or are multiplied by a constant,
\item operators with both the components either holomorphic or anti-holomorphic, as for example $J_{i+} J_{j+}$ $\; \pbar J_{i-}\;,N_{0-}N_{0-}$, come with a space-time dependence given by $\frac{v}{\vbar}$ and ${\vbar\over v}$, (\emph{finite terms}).
\end{itemize}

Here we show the explicit cancellation of the logarithmic divergent terms. Notice that, since the logarithmic terms are independent of the adopted regularization procedure, they have to cancel in any scheme we choose \cite{Mikhailov:2007mr}. One would like to see the same cancellation for the second type of terms (finite terms), however they seem to be really subtle. We leave the analysis of this second type of terms for future investigations.

\subsubsection{Matter currents}

For the quantum fields $X$ the free propagators $A^{-1}$ in \eqref{prop-correction} are
\footnote{The coefficient for the propagator is fixed by $\p\pbar\log|z|^2=2\pi\dl^{(2)}(z)$ and the $\dl$-function in the complex plane is normalized as in \cite{Polchinski:1998rq}.}
\be
&& \langle  X^\mu(x) X^\nu(y) \rangle =-C^{\mu\nu}\log{\gm|v|^2}\\ \nn
&&\langle X^\al(x) X^{\btd}(y) \rangle =-C^{\al{\btd}}\log{\gm|v|^2}\\ \nn
&&\langle X^{\ald}(x) X^\bt(y)  \rangle =-C^{\ald\bt}\log{\gm|v|^2}\\ \nn
\ee
with $|v|=|x-y|$ and $\gm$ is the IR cut-off. $C^{\mu\nu}$ and $C^{\al\btd}=-C^{\btd\al}$ are the inverse of the invariant tensors $C_{\mu\nu}$, $C_{\al\ald}$ and $C_{\ald\al}$, see \appref{app:notation} for details.

At the leading order and from the expansion \eqref{Jmatter-exp}, the general OPE for the matter currents is \cite{Puletti:2006vb}:
\be
\label{opej}
&& J_+^A(x)J_-^B(y) \cong\\ \nn
&&\cong \langle \widetilde{J}_+^A(x)\widetilde{J}_-^B(y)\rangle + {1\over R^2}\Big( \langle\p X^A(x) \pbar X^B(y)\rangle+
\langle \p X^A(x)[\widetilde{J}_-,X]^B(y)\rangle \\ \nn
&&+ \langle[\widetilde{J}_+,X]^A(x)\; \pbar X^B (y) \rangle
+ \langle[\widetilde{J}_+,X]^A(x) [\widetilde{J}_-,X]^B (y)\rangle\Big)+...\;.
\ee
We will not consider the classical contribution $\langle\widetilde{J}_+^A(x)\widetilde{J}_{-}^B(y)\rangle$. $J_+^A(x)J_-^B(y)$ is indeed computed from all the possible contractions of the quantum fields $X$ from $\langle J_{+}^{(1)}(x)J_{-}^{(1)}(y)\rangle$.

The results for the matter OPE's are listed in \appref{app:OPE-results}.


\subsubsection{Matter-Ghost currents}

From the expansions \eqref{Jmatter-exp} and \eqref{N-exp} it follows the expression for the OPE between ghost and matter currents, e.g.

\be
   \label{ope-nj}
   &&\Np(x)J_{i-}(y)\cong {1\over R^2} \big( \langle\{\omp,\tilde{\lm}_3\}(x)\,\pbar X_i(y) \rangle
     +\langle \{\tilde{\om}_{1+},\lm_3\}(x)\,\pbar X_i(y) \rangle\big)+... \\ \nn
   &&\Nm(x)J_{i+}(y)\cong {1\over R^2} \big( \langle \{\omm,\tilde{\lm}_1\}(x)\,\p X_i(y) \rangle +
   \langle\{\tilde{\om}_{3-},\lm_1\}(x)\,\p X_i(y) \rangle\big)+... \;,\\ \nn
\ee
where i=1,2,3.
The only possibility to couple matter fields $X$ and ghost fields $\lm,\;\om$ is through the vertex in \eqref{4-leg-vertex-matter-ghost}, see \figref{fig:tree-vertex-gm}. Such a vertex contains already a classical matter current and a classical ghost field ($\lm$ or its conjugate $\om$). This means that the contraction is already at order $\sim J^2$ and it will produce only logarithmic terms.

By dimensional analysis there is no OPE between the ghost currents and the gauge field $J_0$ at this order because it will involve at least the insertion of three classical currents. Again the results are in \appref{app:OPE-results}.

\subsubsection{Ghost-Ghost currents}

From \eqref{N-exp} the OPE's are

\be
\label{ope-nn}
  && \Np(x)\Nm(y) \cong
  \\ \nn
  &&\cong{1\over R^2}\big(\langle\{\omm,\tilde{\lm}_1\}(x)\{\omp,\tilde{\lm}_3\}(y)\rangle
                                         + \langle\{\omm,\tilde{\lm}_1\}(x)\{\tilde{\om}_{1+},\lm_3\}(y)\rangle\\ \nn
                                      &&+\langle\{\tilde{\om}_{3-},\lm_1\}(x)\{\omp,\tilde{\lm}_3\}(y)\rangle
                                       +\langle\{\tilde{\om}_{3-},\lm_1\}(x)\{\tilde{\om}_{1+},\lm_3\}(y)\rangle\big)
                                       +...\;.
\ee
The ghost fields can be contracted using the interaction terms in \eqref{eff-action-gg}, which contain vertices with two classical ghosts ($\lm,\om$), see \figref{fig:tree-vertex-gg}. From the OPE \eqref{ope-nn} we have two external (classical) legs, this implies again that it can produce at least dimension-two operators and they will be logarithmic divergent. The results are in \appref{app:OPE-results}.


\subsection{Normal order: general structure}
\label{sec-normal-ord}

At this order ${1\over R^2}$ the currents might get renormalized, thus if one wants to control the divergences in $\F$ \eqref{def-f} and in the variation of $\Om$ \eqref{monodromy}, one needs to take into account also the internal contractions in the currents. Explicitly this implies that we have to consider the contractions on the same point for the quantum fluctuations contained in $J$, i.e.
\[
\langle J_{\pm}^{(2)}(x)\rangle = \half \langle[\partial_{\pm} X,X](x)\rangle +\half
 \langle [[\widetilde J_{\pm},X],X](x)\rangle
\]

\begin{figure}
\begin{center}
\includegraphics[scale=1]{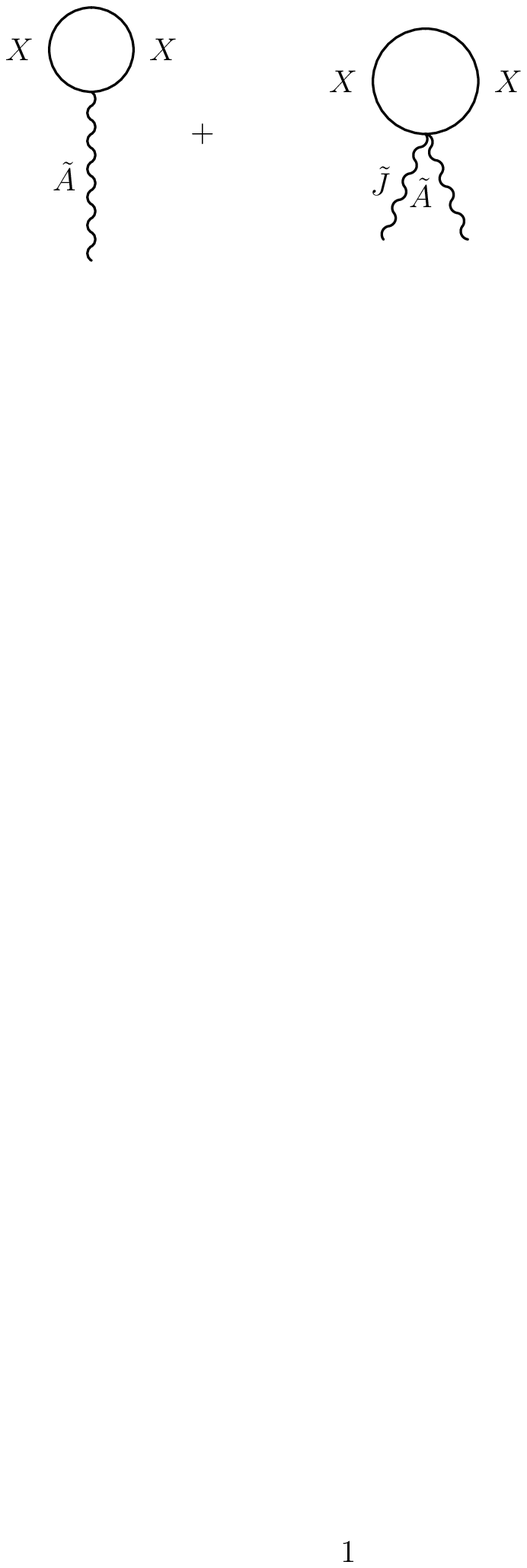}
\end{center}
  \caption{\textbf{Normal ordering diagrams.} The classical matter current is labelled by $\;\tilde{J}\;$, while $\tilde A$ represents the (classical) ghost/matter current. The fields $X$ are the quantum fluctuations.}
\label{fig:loop-diag-mm}
\end{figure}

These loop diagrams are important in order to cancel the divergences coming from the Wilson expansion of the currents. Notice that in the first diagram in \figref{fig:loop-diag-mm} the classical current can be Taylor-expanded producing a dimension-two operator. However in the effective computation we need only the first diagram, since in \eqref{def-no-comm} such a diagram is already multiplied by a classical current.

Thus the normal ordering prescription consists in computing all the contractions in the same point%
\footnote{
In principle we should consider also the internal contractions for the ghost currents. However they do not contribute to any logarithmic divergences but only to finite terms.
}.
This means to consider all tadpole and self-energy diagrams that can be present at order $1/R^2$.

%
%


\section{Field strength}
\label{sec:field-strength}

In the next subsections we explain how we proceed for the computations of the field strength and we prove that all the UV divergences cancel. We performed the entire and complete analysis of $\F$, however in \secref{sec:commutator} we provide two examples for the sectors $z^{1/2}$ and $z^{-1}$, and in \appref{app:comm-results} we report other three examples, i.e. $z^0$, $z^2$ and $z^{3/2}$. These cases are indeed enough to recover the full field strength thanks to the exact symmetry \eqref{symm-f}.

\subsection{Strategy}
\label{sec:strategy}

We have seen that any current (matter and ghost) is expanded around a classical solution, consequently since $\Jflat$ is given by \eqref{def-flat-connection} it becomes
\[
\label{exp-conn}
\Jflat_{\pm}\rightarrow \tilde\Jflat_{\pm}+ {1\over R}\Jflat_{\pm}^{(1)} +{1\over R^2}\Jflat_{\pm}^{(2)}\;,
\]
and analogously for the field strength
\[
\label{exp-f}
\F_{+-} \rightarrow \tilde\F_{+-} + {1\over R}\F_{+-}^{(1)} +{1\over R^2}\F_{+-}^{(2)}\;,
\]
with $\tilde\F_{+-}=0$.

We want to investigate $\F_{+-}^{(2)}$ and show that it is not affected by any logarithmic divergence such that also the variation of the monodromy matrix $\Om$ does not contain UV divergent terms%
\footnote{
In principle we could expect interactions between the field strength and the connections contained in the path ordered exponential in \eqref{var-coord-mm}. However at this order and up to dimension-two operators all the operators contained in $\F$ and inserted in the modified contour interact between themselves and produce just classical currents, which cannot interact with anything else since they satisfy the classical equations of motion \eqref{eom-matter}, \eqref{eom-ghost} and the field strength is zero classically \eqref{zero-curv-eq}. Indeed expanding $\F$ and $\Om$ as in \eqref{exp-conn} and in \eqref{exp-f}, then there might be possibilities of interaction in ${1\over R}\F^{(1)} ({1\over R}\int\Jflat^{(1)} +{1\over R}\int\Jflat^{(1)}\int\widetilde\Jflat)$  and in ${1\over R^2}\F^{(2)}\int\widetilde\Jflat\widetilde\Jflat$. The first term vanishes by using the equations of motion, as can be directly checked. The second term is possible if ${1\over R^2}\F^{(2)}$ contains dimension-zero operators, which is not the case due to the antisymmetry of $\F$. }.
One can write the curvature tensor as
\[
\label{normal-order}
\mathcal{F}_{+-}(z)\;=\;
:\mathcal{F}_{+-}(z):
+ \sum_k C_k (\eps)  \mathcal{O}_{k;+-}(z) \,.
\]
The symbol $:\;:$ denotes the normal ordering prescription, namely the contribution to $\F$ coming from the internal contractions in the currents, while the sum $\sum_k C_k (\eps) \mathcal{O}_k$ is the operator product expansion (OPE) which, by definition, takes into account the effects of the  operator $\Jflat\Jflat$.
Explicitly:
\[
\partial_+ \mathcal{J}_--\partial_- \mathcal{J}_+ \;= \;
:\partial_+ \mathcal{J}_--\partial_- \mathcal{J}_+\;:
\]
and
\be
\label{comm-J-flat}
[\mathcal{J}_+(x),\mathcal{J}_-(y)] \;=\;
:[\mathcal{J}_+(x),\mathcal{J}_-(y)]:+
f_{BC}^A \,\Jflat_+^B(x) \Jflat_-^C(y)\,t_A \;=\nn \\
=:\,[\mathcal{J}_+(x),\mathcal{J}_-(y)]\,:
+ \sum_k C_k (\eps)  \mathcal{O}_{k;+-} (\si) \;\;,
\ee
when $x-y\sim \eps$ and $\si\equiv\frac{x+y}{2}$.

In particular we are mainly interested in the commutators contained in $\F$ because $[\Jflat_+,\Jflat_-]$ naturally contains the possible dangerous short-distance interactions between the currents \cite{Abdalla:1980jt}.

Since $\Jflat$ and $\F$ are linear combinations of matter and ghost currents and $z$ is the coefficient of such combinations, this implies that one has a set of independent equations to verify at leading order, because obviously the cancellation of divergences must be independent of the values of $z$. For the commutators we list for completeness the equations we need to compute
\be
\label{comm-explicit-list}
&&[\Jflat_+,\Jflat_-]= \\ \nn
   && = [\Jop,\Jom]-[\Jop,\Nm]-[\Np,\Jom] + 2[\Np,\Nm]\\ \nn
    &&+[\Jtwop,\Jtwom]+[\Jthreep,\Jonem]+[\Jonep,\Jthreem]\\ \nn
    &&+ z^{-2}\big([\Jop,\Nm]-[\Np,\Nm]\big)\\ \nn
    &&+z^{2}\big([\Np,\Jom]-[\Np,\Nm]\big)\\ \nn
    &&+z^{-1}\big([\Jop,\Jtwom]+[\Jthreep,\Jthreem]+[\Jtwop,\Nm]-[\Np,\Jtwom]\big)\\ \nn
    &&+z^{-3/2}\big([\Jop,\Jthreem]+ [\Jthreep,\Nm]-[\Np,\Jthreem]\big)\\ \nn
    &&+z^{-1/2}\big([\Jop,\Jonem]+[\Jtwop,\Jthreem]+[\Jthreep,\Jtwom]+[\Jonep,\Nm]-[\Np,\Jonem]\big)\\ \nn
    &&+z\big([\Jtwop,\Jom]+[\Jonep,\Jonem]-[\Jtwop,\Nm]+[\Np,\Jtwom]\big)\\ \nn
    &&+z^{1/2}\big([\Jthreep,\Jom]+[\Jtwop,\Jonem]+[\Jonep,\Jtwom]-[\Jthreep,\Nm]+[\Np,\Jthreem]\big)\\ \nn
    &&+z^{3/2}\big([\Jonep,\Jom]-[\Jonep,\Nm]+[\Np,\Jonem]\big)\;.
\ee
We refer to each particular combination labelled by $z^s$ as a sector, since eventually the different powers of the spectral parameter distinguish the different subalgebras.

The strategy is to calculate the contributions to the commutators in $\F$ from the OPE's and from the internal contractions separately following \eqref{comm-J-flat} and show that indeed the divergent terms cancel against each other. Each commutator in \eqref{comm-explicit-list} will be written as
\be
&&[J_+ (x),J_- (y)]^A\; =\; f^A_{BC}\; J_+^B (x) J_-^C (y) + :[J_+ (x),J_- (y)]^A:\;,\\ \nn
&&[J_+(x), N_-(y)]^A =f_{B[\mu\nu]}^A J_+^B(x)N_-^{[\mu\nu]}(y)+ :[J_+(x), N_-(y)]^A:\;,\\ \nn
&&[J_-(x), N_+(y)]^A=f_{B[\mu\nu]}^A J_-^B(x)N_+^{[\mu\nu]}(y)+ :[J_-(x), N_+(y)]^A:\;,\\ \nn
&&[N_+(x),N_-(y)]^{[\mu\nu]}= f^{[\mu\nu]}_{[\lm_1\rho_1][\lm_2\rho_2]}N^{[\lm_1\rho_1]}_+(x) N^{[\lm_2\rho_2]}_-(y)
                           +:[N_+(x),N_-(y)]^{[\mu\nu]}:\;,
\ee
where again all the first terms are computed from the OPE's while the second is the normal ordered commutator%
\footnote{Notice that there will be no contribution to the logarithms from the internal contractions of the ghost currents.}.
For the OPE contribution this means to compute the expressions \eqref{opej}, \eqref{ope-nj}, \eqref{ope-nn}, while the normal-ordering contributions are given for example by
\[
\label{def-no-comm}
:[J_+ (x), J_-(y)]:\;
= {1\over R^2}\,[\langle J_+^{(2)}(x)\rangle, \widetilde J_-(y)] + {1\over R^2}\, [\widetilde J_+(x), \langle J_-^{(2)}(y)\rangle]\;+....
\]

As explained above there are in principle other contributions to the normal ordered $\F$ coming from
\[
\label{F-derivatives}
:\;\p\Jflat_--\pbar\Jflat_+:\;=\;{1\over R^2} \,\big(\,\p \langle \Jflat^{(2)}_-\rangle - \pbar \langle \Jflat_+^{(2)}\rangle\,\big)\;+....
\]
It turns out that these terms do not contribute, as one could expect naively. This does not mean that $\langle\Jflat^{(2)}\rangle$ vanishes, however it turns out that at this order the only non-vanishing contributions from $ \langle J^{(2)}\rangle$ have already dimension two. Consequently when the derivatives act on it (e.g. $\p \langle J^{(2)}_-\rangle$), they can act only on the propagators, and make the expression \eqref{F-derivatives} vanish due to the translational invariance
\footnote{
There is however an exception for the sector labelled by $z^0$ in $\mathfrak g_0$. The key point in this case is that we are computing $\langle J_{0\pm}^{(2)}\rangle$ which has non-vanishing terms of dimension one, consequently now the derivatives can act on the current itself giving rise to a dimension-two operator $\partial J$. More details are in \appref{app:comm-results}.
}.
%


\subsection{Some specific examples}
\label{sec:commutator}


We want to underline some common features to all the sectors $z^s$ of \eqref{comm-explicit-list}.

For the OPE contributions many terms are generated and they cancel against each other after manipulating them by means of the graded Jacobi identities. What remains eventually is always \emph{only one} logarithmic divergent term, this term is always a mixed commutator between ghost and matter currents. It might seem strange that the symmetry between the two Lorentz currents $N_{0+}$ and $N_{0-}$ is broken.
However this term is exactly balanced by the other logarithm produced in the internal contractions.

As we said $:[\Jflat_+, \Jflat_+]:$ is a linear combination of commutators such as \eqref{def-no-comm}. Once we insert the current expansions \eqref{Jmatter-exp} and \eqref{Jzero-exp} in the above expression \eqref{def-no-comm} and contract the fields, many terms are produced, but again it remains \emph{only one} which is logarithmic divergent.
It always comes from the internal contractions for the gauge fields, i.e. its origin is in terms as $[\langle J_{0+}^{(2)}\rangle, \widetilde J_{i-}]$ or $[\widetilde J_{i+}, \langle J_{0-}^{(2)}\rangle]$.
In particular looking at the gauge field expansion \eqref{Jzero-exp}, the terms
\[
{1\over 2R^2}\,\langle\, [\partial_{\pm} X_1,X_3]+[\partial_{\pm} X_3,X_1]+[\partial_{\pm} X_2,X_2]\,\rangle
\]
can be contracted using the vertices $N_{\pm}[\partial_{\mp} X,X]$ in \eqref{three-leg-gm}, giving rise in this way to a mixed and logarithmic divergent commutator between ghost and matter currents, which is exactly what we need in order to cancel the divergent term coming from the OPE's.
The full calculation is in the next subsections, we will just keep track of the logarithmic terms.


\subsubsection{Example 1: commutators of $z^{-1}$}
\label{subsec:A2-commutator}

The contributions to $[\Jflat_+,\Jflat_-]^A$ for $A=2$ is split in two sets with respect to the spectral parameter \eqref{comm-explicit-list}, in this section we consider the commutators with the coefficient $z^{-1}$, i.e.
\be
\label{comm-mu-1}
[\Jthreep,\Jthreem]
+[J_{0+},\Jtwom]
-[N_{0+},\Jtwom]
-[\Nm,J_{2+}]\; .
\ee


\paragraph{OPE's.}

For $z^{-1}$  we need to sum the OPE's corresponding to the commutators in \eqref{comm-mu-1}, i.e. \eqref{opej-j3j3}, \eqref{ope-j0j2}, \eqref{ope-j2nbar} and \eqref{ope-j2barn}. For practical reason it is convenient to collect the terms for the different types of dimension-two operators.
\begin{itemize}
\item The derivatives come from $[J_{3+},J_{3-}]$ \eqref{opej-j3j3} and from $[J_{0+},J_{2-}]$ \eqref{ope-j0j2} and after manipulating the structure constants one obtains
\be
&&   f^{\lm}_{[\mu\nu]\rho}f^{\rho[\mu\nu]}_\si \big(\p J^\si_+ \frac{v}{2\bar v}+\half \pbar J^\si_+\big)\cr
&&    +f^\lm_{\al\bt}f^{\al\bt}_\si \big(\p J^\si_+ \frac{v}{2\bar v}+\half \pbar J^\si_+ -\half \p J^\si_-\log \gm|v|^2\big)= \cr
&&   -\half f^\lm_{\al\bt}f^{\al\bt}_\si \p J_-^\si\log \gm|v|^2 \;.
\ee
\item For the matter currents the contribution is only from $[J_{0+},J_{2-}]$ \eqref{ope-j0j2}, since the other possible term in $[J_{3+}, J_{3-}]$ \eqref{opej-j3j3} is antisymmetric in $\al\;,\bt$
\be
&&+ f^{\lm}_{[\mu\nu]\rho} f^{[\mu\nu]}_{\al\ald}f^{\ald \rho}_\bt J^\al_+ J_+^\bt\frac{v}{\bar v}
- f^{\lm}_{[\mu\nu]\rho} f^{[\mu\nu]}_{\al\ald}f^{\ald \rho}_\bt J_+^\al J_-^\bt\log\gm|v|^2 =\\ \nn
&& = -\half f^\lm_{\ald\btd}f^{\ald\btd}_\rho f^\rho_{\al\bt} J_+^\al J_-^\bt\log \gm|v|^2
\ee
\item There are contributions from all the OPE's for the commutator formed by matter and ghost currents.
We have from $[J_{3+},J_{3-}]$ \eqref{opej-j3j3}
\be
- f^\lm_{\al\bt}f^{\bt\gm}_\si f^{\al}_{\gm[\mu\nu]}J_-^\si N^{[\mu\nu]_+}\log \gm|v|^2
+ f^\lm_{\al\bt}f^{\al}_{\si\ald}f^{\ald\bt}_{[\mu\nu]} J^\si_+
N_-^{[\mu\nu]}\log \gm|v|^2 
+...\;,
\ee
from $[J_{0+},J_{2-}]$ \eqref{ope-j0j2}
\[
f^\lm_{[\mu_1\nu_1]\rho_1} f^{[\mu_1\nu_1]}_{\si\rho_2}f^{\rho_1\rho_2}_{[\mu_2\nu_2]}
       J^\si_+
       N_-^{[\mu_2\nu_2]}\log \gm|v|^2
       +...\;,
\]
from $[J_{2+},\Nm]$ \eqref{ope-j2nbar}
\[
-f^\lm_{[\mu_1\nu_1]\rho} f^{\rho}_{\si[\mu_2\nu_2]}f^{[\mu_1\nu_1][\mu_2\nu_2]}_{[\mu_3\nu_3]} N_+^{[\mu_3\nu_3]}J_-^\si \log \gm|v|^2\;,
\]
and from $[N_{0+},J_{2-}]$ \eqref{ope-j2barn}
\[
-f^\lm_{[\mu_1\nu_1]\rho} f^{\rho}_{\si[\mu_2\nu_2]}f^{[\mu_1\nu_1][\mu_2\nu_2]}_{[\mu_3\nu_3]} N_-^{[\mu_3\nu_3]}J_+^\si \log \gm|v|^2\;.
\]
One can already see that due to the commutator containing the gauge field $[J_{0+},J_{2-}]$, there is no symmetry between the two ghost currents.
Manipulating the structure constants with the graded Jacobi identities and taking into account only the logarithmic terms one has
\be
&&-\half f^\lm_{\al\bt} f^{\al\bt}_\rho f^\rho_{\si[\mu\nu]}N_-^{[\mu\nu]}J^\si_+ \log \gm|v|^2 \\ \nn
&&+ \half(-f^\lm_{\al\bt} f^{\al\bt}_\rho f^\rho_{\si[\mu_2\nu_2]}
   + f^{[\mu_1\nu_1][\mu_3\nu_3]}_{[\mu_2\nu_2]}f^{[\mu_4\nu_4]}_{[\mu_3\nu_3][\mu_1\nu_1]}f^\lm_{\si[\mu_4\nu_4]})
   J^\si_- N_+^{[\mu_2\nu_2]}\log\gm|v|^2 \;.
\ee
\end{itemize}
Collecting all the contributions the result for \eqref{comm-mu-1} is
\be
\label{result-comm-mu-1}
&&-\half f^\lm_{\al\bt}f^{\al\bt}_\si \log \gm|v|^2
  \big( \p J_-^\si
        +f^\si_{\al\bt} J_+^\al J_-^\bt
        +f^\si_{\rho[\mu\nu]} N_-^{[\mu\nu]}J^\rho_+
        +f^\si_{\rho[\mu\nu]}J_-^\si N_+^{[\mu\nu]}\big) \\ \nn
&&+\half f^{[\mu_1\nu_1][\mu_2\nu_2]}_{[\mu_4\nu_4]}f^{[\mu_3\nu_3]}_{[\mu_2\nu_2][\mu_1\nu_1]}f^\lm_{\si[\mu_3\nu_3]}
        J^\si_- N_+^{[\mu_4\nu_4]}\log\gm|v|^2 =\\ \nn
&&= \half f^{[\mu_1\nu_1][\mu_2\nu_2]}_{[\mu_4\nu_4]}f^{[\mu_3\nu_3]}_{[\mu_2\nu_2][\mu_1\nu_1]}f^\lm_{\si[\mu_3\nu_3]}
        J^\si_- N_+^{[\mu_4\nu_4]}\log\gm|v|^2 \;,
\ee
since the first line is zero by classical equations of motion \eqref{eom-matter}.


\paragraph{Internal contractions.}

Inserting the expansion \eqref{Jmatter-exp}, \eqref{Jzero-exp} in the commutators and contracting the quantum fields many terms vanish because the contraction of two structure constants is through one fermionic and one bosonic index. Basically when there is an odd number of fermions in the same double commutator containing the fields to be contracted, the result is zero. Using the algebraic identity \eqref{id2} in \appref{app:alg-identities}, which is nothing but the vanishing dual Coxeter number \cite{Vallilo:2002mh, Berkovits:1999zq}, the only non-vanishing contribution is from $[J_{0+},\Jtwom]$
\be
\label{ope-j0jbar2-samep}
:[J_{0+},\Jtwom]:\, =
\half [\,\langle\,[\p X_1,X_3]+[\p X_3,X_1]+[\p X_2,X_2]\,\rangle\,,\Jtwom]+...=\cr
=\half f^\lm_{[\mu_1\nu_1]\si} f^{[\mu_1\nu_1]}_{[\mu_3\nu_3][\mu_4\nu_4]}f^{[\mu_4\nu_4][\mu_3\nu_3]}_{[\mu_2\nu_2]}
     J_-^\si N_+^{[\mu_2\nu_2]}\log\gm|v|^2\;+...,
\ee
One can see directly that this term \eqref{ope-j0jbar2-samep} cancels exactly the divergence coming from the result \eqref{result-comm-mu-1} for the OPE contributions, leaving the sector $z^{-1}$ free of UV divergences.


\subsubsection{Example 2: commutators of $z^{1/2}$ }
\label{subsec:A3-commutator}

The second example which we want to treat explicitly is for the value of the spectral parameter $z^{1/2}$, i.e.
\[
\label{mu1-2}
[J_{3+},\Jom]-[J_{3+},\Nm]+[\Np,\Jthreem]+[J_{1+},\Jtwom]+[J_{2+},\Jonem]\;.
\]


\paragraph{OPE's.}
We need to collect the OPE's \eqref{ope-j3jbar0}, \eqref{ope-j3nbar}, \eqref{ope-jbar3n}, \eqref{ope-j1jbar2} and \eqref{ope-j2jbar1}, in order to compute the expression \eqref{mu1-2}.

\begin{itemize}
\item For the derivatives which are contained in $[J_{3+},\Jom]$ \eqref{ope-j3jbar0}, $[J_{1+},\Jtwom]$ \eqref{ope-j1jbar2} and in $[J_{2+},\Jonem]$ \eqref{ope-j2jbar1}, there is no contribution due to the algebraic identities \eqref{id1}.
\item The contributions to the commutator between two matter currents are from $[J_{1+},\Jtwom]+[J_{2+},\Jonem]$, \eqref{ope-j1jbar2}, \eqref{ope-j2jbar1}
\be
f^\dl_{\nu\ald}f^\nu_{\gmd\btd}f^{\ald\gmd}_\mu \big(J_-^{\btd}J_+^\mu\log\gm|v|^2
                                               +J_+^{\btd}J_-^\mu\log\gm|v|^2
                                              \big)\;,
\ee
and from $[J_{3+},\Jom]$ \eqref{ope-j3jbar0} with
\be
 &&-f^\dl_{\al[\mu\nu]}f^{[\mu\nu]}_{\mu\nu}f^{\nu\al}_{\btd} 
                                                J_+^{\btd}J_-^\mu \log\gm|v|^2
+f^\dl_{\al[\mu\nu]}f^{[\mu\nu]}_{\btd\gm}f^{\gm\al}_\mu
                                                   J_+^\mu J_-^{\btd} \log\gm|v|^2
                                                   +...\;.
\ee
Manipulating the structure constants and summing the two contributions above, the terms turn out to be free from logarithms.%
%
\item For the mixed commutator, $[J_{1+},\Jtwom]$ and $[J_{2+},\Jonem]$ cancel against each other all the logarithmic terms containing ghost Lorentz currents and matter currents in \eqref{ope-j1jbar2} and in \eqref{ope-j2jbar1}. Thus from \eqref{ope-j1jbar2} and \eqref{ope-j2jbar1} there is no contribution. The only OPE's which contribute are $[J_{3+},\Jom]$ \eqref{ope-j3jbar0} with
\be
&&f^\dl_{\al[\mu_1\nu_1]}f^{[\mu_1\nu_1]}_{\bt\gmd}f^{\gmd\al}_{[\mu_2\nu_2]}
  J_-^\bt  N_+^{[\mu_2\nu_2]}\log\gm|v|^2\\ \nn
&&=\half f^{[\mu_3\nu_3][\mu_4\nu_4]}_{[\mu_2\nu_2]}f^{[\mu_1\nu_1]}_{[\mu_3\nu_3][\mu_4\nu_4]}f^\dl_{[\mu_1\nu_1]\bt}
  J_-^\bt N^{[\mu_2\nu_2]}_+ \log\gm|v|^2
  \;+...,
\ee
and $[J_{3+},\Nm]$ \eqref{ope-j3nbar}, $[\Np,\Jthreep]$ \eqref{ope-jbar3n} with
\[
-\half f^{[\mu_3\nu_3][\mu_4\nu_4]}_{[\mu_2\nu_2]}f^{[\mu_1\nu_1]}_{[\mu_3\nu_3][\mu_4\nu_4]}f^\dl_{[\mu_1\nu_1]\bt}
\big( J_-^\bt N_+^{[\mu_2\nu_2]}+ J_+^\bt N_-^{[\mu_2\nu_2]}\big)\log\gm|v|^2\;.
\]
Thus one obtains for the OPE contributions in $z^{1/2}$
\[
\label{result-mu1-2}
-\half f^{[\mu_3\nu_3][\mu_4\nu_4]}_{[\mu_2\nu_2]} f^{[\mu_1\nu_1]}_{[\mu_3\nu_3][\mu_4\nu_4]} f^\dl_{[\mu_1\nu_1]\bt}
 J_+^\bt N_-^{[\mu_2\nu_2]}\log\gm|v|^2\;,
\]
where we have used the identities in \appref{app:alg-identities}.
\end{itemize}

\paragraph{Internal contractions.}

Since the expansions \eqref{Jmatter-exp} for $J_3$ and $J_1$ contain an odd number of fermions they do not contribute to the loop diagrams. Thus it follows that the only possible contributions to the internal contractions come from $J_0$ and $J_2$ contained in \eqref{mu1-2}. However also in this case it turns out that all the logarithms in $J_2$ cancel due to the identities \eqref{id2}. Again the commutator producing the logarithmic term is the one involving the gauge field, i.e. $:[J_{3+},\Jom]:$ with
\be
\label{ope-j3jbar0-samep}
\half\big([J_{3+},\,\langle\,[\pbar X_3,X_1]+ [\pbar X_3,X_1]+[\pbar X_2,X_2]\,\rangle\,]\big)=\cr
= \half f^\dl_{\al[\mu_4\nu_4]}f^{[\mu_4\nu_4]}_{[\mu_1\nu_1][\mu_2\nu_2]}f_{[\mu_3\nu_3]}^{[\mu_2\nu_2][\mu_1\nu_1]}
        J_+^\al N_-^{[\mu_3\nu_3]}\log\gm|v|^2\;+....
\ee
which matches perfectly the logarithmic commutator in \eqref{result-mu1-2}.


\section{Conclusions}
\label{sec:conclusions}

In this work we have directly proved that the monodromy matrix $\Om$ \eqref{monodromy} of the pure spinor type IIB superstring on $AdS_5\times S^5$ is independent of path deformations at first order in perturbation theory. Indeed an anomaly in the variation of the monodromy matrix is equivalent to the insertion of a local ghost-number zero and conformal dimension $(1,1)$ operator $\mathcal O^{(1,1)}$, whose ansatz is expressed in \eqref{O-def}. Demanding that $\mathcal O^{(1,1)}$ satisfies the equation \eqref{brst}, rules out the existence of such an operator and consequently any possible anomaly. Since our arguments are actually valid to all orders in perturbation theory and the independence of the contour for the monodromy matrix $\Om$ leads to the absence of logarithmic divergences \cite{Mikhailov:2007mr}, then this implies that $\Om$ is finite to all orders in $\al'$.

In the second part of the work we have explicitly showed that the field strength $\F$ \eqref{def-f} is UV-finite at the leading order. Notice that $\F$ obeys to the equation \eqref{brst} and classically vanishes. In order to calculate $\F$ we need to know the short-distance behavior for the currents, namely their OPE's. This is done by means of the background field method, expanding the currents around a classical solution and computing the effective action for the quantum fluctuations. We have reported two examples in the main text for the commutators labelled by $z^{-1}$ and $z^{1/2}$ in order to visualize how the log-divergences cancel. The main point is that in the OPE's of the commutators after some algebraic manipulations, only one logarithmic divergent term eventually remains which is exactly cancelled by a term coming from the normal ordered commutators.

The two complementary arguments provide a direct check of the integrability of the pure spinor superstring in $AdS$ space at the leading order and they strongly suggest that it should be integrable at all orders in perturbation theory.

\bigskip
\subsection*{Acknowledgments}
\bigskip

I would like to thank K. Zarembo for suggesting the problem and for his constant guidance. I am grateful to A. Mikhailov for the initial collaboration on this project and for many helpful discussions. I also benefited from the initial collaboration with S. Schafer-Nameki and from useful comments by G. Grignani and O. Ohlsson Sax. I thank VR for partial financial support and Universit\`a di Perugia for the hospitality during the last part of the work. Finally I thank K. Zarembo, A. Mikhailov and N. Berkovits for reading the manuscript and for their valuable comments.


\appendix

\section{Notation and conventions}
\label{app:notation}

The $\mathfrak{psu}(2,2|4)$ Lie-algebra  has a $\bold{Z}_4$ inner symmetry \cite{Berkovits:1999zq}, which decomposes it in
\[
\mathfrak{psu}(2,2|4)\equiv\mathfrak{g} = \mathfrak{g}_0 +\mathfrak{g}_1+\mathfrak{g}_2+\mathfrak{g}_3\;.
\]
$\mathfrak{g}_0 +\mathfrak{g}_2$ are the bosonic subalgebras, in particular $\mathfrak{g}_0$ is the $\bold{Z}_4$-invariant subalgebra for the gauge group $SO(4,1)\times SO(5)$, $\mathfrak{g}_2$ contains the remaining bosonic elements, while the fermionic subalgebras are $\mathfrak{g}_1+\mathfrak{g}_3$.
Hence the $\mathfrak{psu}(2,2|4)$ generators are
\[
t_A=\{t_{[\mu\nu]}^0,\, t_\mu^2\,, t_{\ald}^1\,,t_\al^3\}\;.
\]
The indices labelling the tangent space are $A=(\mu, [\mu\nu], \ald,\al)$, with $\mu\in \mathfrak{g}_2\, ,\mu=0...9$, $[\mu\nu]\in\mathfrak{g}_0\,,[\mu\nu]=-[\nu\mu]$, and the fermionic indices $\al\in\mathfrak{g}_3\,,\al=1...16$, $\ald\in\mathfrak{g}_1\,,\ald=1...16$ label the two sixteen-component Majorana-Weyl spinors in ten dimensions.\\
The ten-dimensional Dirac matrices $\Gamma^\mu$ are real and symmetric and in the reducible Mayorana-Weyl representation they consist of two symmetric $16 \times 16$ matrices $\gm^\mu$ in the off-diagonal.

The $\bold{Z}_4$-grading respects the structure of the algebra (i.e. $[\mathfrak{g}_m,\mathfrak{g}_n]\in \mathfrak{g}_{m+n\; (\mod 4)}$) and the invariant bilinear form on $\mathfrak{psu}(2,2|4)$. Such invariant form is defined in terms of the super-trace $\textrm{Str}$ in the fundamental representation  and the fact that is $\bold{Z}_4$-invariant means that $\textrm{Str}(t_A t_B)=0$ unless $A+B=0 \;(\mod 4)$.
The super-trace is cyclic up to a minus sign for the exchange of fermions, and in particular
\be
\textrm{Str}(t_\mu^2 \;t_\nu^2)= \textrm{Str}( t_\nu^2 \;t_\mu^2)=C_{\mu\nu}
\nn \\
\textrm{Str}(t_{\ald}^1 \;t_{\bt}^3)=-\textrm{Str}( t_{\bt}^3\; t_{\ald}^1)= C_{\ald\bt}
\nn \\
\textrm{Str}(t_{\al}^3\; t_{\btd}^1)=-\textrm{Str}( t_{\btd}^1 \;t_{\al}^3)= C_{\al\btd}
\ee

For brevity we will use also the convention that $i=1,2,3$ labels the algebraic indices corresponding to $\mathfrak{g}_1, \mathfrak{g}_2$ and $\mathfrak{g}_3$ respectively, i.e. $J_i\equiv J_{|\mathfrak{g}_i}$, while the lower index $0$ labels the projection to the gauge algebra $\mathfrak{g}_0$, i.e. $J_0\equiv J_{|\mathfrak{g}_0}$.

Furthermore the inverse invariant tensors are defined such that
\be
C^{\al\btd}C_{\btd\bt}=\dl^{\al}_{\bt}
\quad
C^{\ald\bt}=-C^{\bt\ald}
\quad
C^{\mu\nu}C_{\nu\lm}=\dl^\mu_\lm
\ee
and we raise the indices in the structure constants according to $f_{AB}^C g^{BD}\equiv f_A^{DC}$, explicitly
\be
f_{\al\bt}^\mu C^{\bt\ald}\equiv f_\al^{\ald \mu}\;,
\qquad
f_{\al \nu}^{\btd}C^{\nu\mu}\equiv f_\al^{\mu\btd}\;.
\ee

\paragraph{Conventions about derivatives and coordinates.}
The world-sheet action is Euclidean and coordinates and derivatives are defined in the following way
\[
z^+\equiv z=x^1+\imath x^2,\qquad z^-\equiv \bar{z}=x^1-\imath x^2\;,
\]
\[
\p\equiv\partial_{z}=\half\big(\partial_1-\imath \partial_2)
\qquad
\pbar\equiv\partial_{\bar z}=\half\big(\partial_1-\imath \partial_2\big)\;,
\]
In a covariant notation the world-sheet indices are $a,b$.


\section{Detailed derivation of equations (\ref{system-A})}
\label{app:brst}

Here we derive the system of linear equations \eqref{system-A} for the unknown functions $A$'s.

We apply the BRST transformations \eqref{brst-transf} to the operator $\mathcal O^{(1,1)}$ defined in \eqref{O-def}. For simplicity we write the results separately for the different terms which form $\mathcal O^{(1,1)}$.
\begin{itemize}
\item Acting with the BRST operator $Q$ on the terms $A^{0+,2-}[\Np,\Jtwom]$$+$$A^{0-,2+}[J_{2+},\Nm]$ one gets
\be
\label{piece:1}
&& Q\cdot(
A^{0+,2-}[\Np,\Jtwom]+
A^{0-,2+}[J_{2+},\Nm])=\\ \nn
&&= A^{0+,2-}\big([[\Np,\Jthreem],\lm_3]+[[\Np,\Jonem],\lm_1]+[\Jonem,[\Np,\lm_1]]\big)\\ \nn
&& + A^{0-,2+}\big([[J_{1+},\Nm],\lm_1]+[[J_{3+},\Nm],\lm_3]+[J_{3+},[\Nm,\lm_3]]\big)\;.
\ee
\item Acting with $Q$ on $A^{2+,3-}[J_{2+},J_{3-}]+A^{0+,3-}[\Np,\Jthreem]$ one obtains
\be
\label{piece:2}
&&Q\cdot (
A^{2+,3-}[J_{2+},J_{3-}]+
A^{0+,3-}[\Np,\Jthreem])=\\ \nn
&& = A^{2+,3-}\big([[J_{3+},\lm_3],\Jthreem]+ [[J_{1+},\lm_1],\Jthreem]
                +[\Jtwop,[\Jtwom,\lm_1]]+[J_{2+},[\Nbar,\lm_3]]\big)\\ \nn
&& + A^{0+,3-}\big([[J_{1+},\lm_3],\Jthreem]+[\Np,[\Jtwom,\lm_1]]+[[\Np,\Nm],\lm_3]\big)\;.
\ee
\item The BRST transformations for the terms $A^{1+,2-}[J_{1+},J_{2-}]+A^{1+,0-}[J_{1+},\Nm]$ provide
\be
\label{piece:3}
&&Q\cdot (
A^{1+,2-}[J_{1+},J_{2-}]+
A^{1+,0-}[J_{1+},\Nm])=\\ \nn
&& = A^{1+,2-}\big([[J_{2+},\lm_3],\Jtwom]+[J_{1+},[\Jthreem,\lm_3]]
              +[J_{1+},[\Jonem,\lm_1]]+[[\Np,\lm_1],\Jtwom]\big)\\ \nn
&& + A^{1+,0-}\big([J_{1+},[\Jthreem,\lm_1]]+[[J_{2+},\lm_3],\Nm]+[\lm_1,[\Np,\Nm]]\big)
\ee
\item When $Q$ acts on the terms $A^{2+,2-}[J_{2+},J_{2-}]+A^{1+,3-}[J_{1+},J_{3-}]$ the result is
\be
\label{piece:4}
&&Q\cdot (
A^{2+,2-}[J_{2+},J_{2-}]+
A^{1+,3-}[J_{1+},J_{3-}])=\\ \nn
&& = A^{1+,3-}\big([[J_{2+},\lm_3],\Jthreem]+[J_{1+},[\Jtwom,\lm_1]]
                               +[[N,\lm_1],\Jthreem]+[J_{1+},[\Nm,\lm_3]]\big)\\ \nn
&& + A^{2+,2-}\big([[J_{3+},\lm_3],\Jtwom]+[[J_{1+},\lm_1],\Jtwom]
                +[J_{2+},[\Jthreem,\lm_3]]+[J_{2+},[\Jonem,\lm_1]]\big)
\ee
\item Finally the BRST transformation of $[\Np,\Nm]$ gives
\be
\label{piece:5}
Q\cdot (A^{0+,0+} [\Np,\Nm])=
A^{0+,0+}([[\Jonep,\lm_3],\Nm]+[\Np,[\Jthreem,\lm_1]])
\ee
\end{itemize}

Considering all the terms above, i.e. \eqref{piece:1, piece:2, piece:3, piece:4, piece:5}, and demanding that the equation \eqref{brst} holds, one obtains for the coefficients $A$'s the following equalities, which can be collected noticing for example that
\begin{itemize}
\item the terms with $[\Np,\Nm]$ are only in \eqref{piece:2} and \eqref{piece:3} and this fixes the first conditions
\[
\label{cond:1}
A^{0+,3-}= A^{1+,0-} = A^{0+,0-}\;;
\]
\item
the terms containing $[\Jonep,\Jthreem]$, namely in \eqref{piece:1} and \eqref{piece:2},
lead to the conditions
\be
A^{1+,0-}= A^{2+,3-} \quad A^{0+,3-}=A^{1+,2-}
\quad A^{1+,3-}=A^{1+,2-}=A^{2+,3-}\;;
\ee
\item
the terms with $[\Jtwom,\Jtwop]$ in \eqref{piece:1} and in \eqref{piece:2} can be rewritten using the equations of motion \eqref{eom-matter}, \eqref{eom-ghost} as
\be
\label{piece:6}
&&A^{2+,3-}[\Jtwop,[\Jtwom,\lm_1]]
+ A^{1+,2-}[[\Jtwop,\lm_3],\Jtwom]=\\ \nn
&&+A^{2+,3-}[\lm_1,[\Jtwom,\Jtwop]]+A^{2+,3-}[\Jtwom,[\Jtwop,\lm_1]]+\\ \nn
&&+A^{1+,2-}[[\Jtwop,\Jtwom],\lm_3]+A^{1+,2-}[[\Jtwom,\lm_3],\Jtwop]
\ee
and they give the following conditions
\be
A^{2+,2-}=A^{2+,3-}=A^{1+,2-}
\qquad
A^{2+,3-}=A^{1+,2-}=0\;.
\ee
\end{itemize}
We need
\[
\label{piece:7}
A^{2+,3-}=A^{1+,2-}=0
\]
in order to cancel $[[\Jtwom,\lm_3],\Jtwop]$ and $[\Jtwom,[\Jtwop,\lm_1]]$ from \eqref{piece:6}, since one cannot use the equations of motion in this case.
These last two conditions \eqref{piece:7} drastically reduce the initial system, because now all the coefficients are related and what remains is only the contribution from \eqref{piece:1}. However since there is no $[N_{0\pm},J_{2\mp}]$ in the r.h.s. of \eqref{piece:1}, this means that there is no solution for the linear combination \eqref{system-A}.


\section{OPE's: results}
\label{app:OPE-results}

Here we report the results for the OPE's up to quadratic terms in the currents. The symbol $\;\widetilde{}\;$ is omitted, however all the currents in the R.H.S. are classical and there is an overall factor $1/R^2$ also omitted.

It is convenient to perform the OPE's in the symmetric point $\si\equiv (x+y)/2$, i.e. $J(x)J(y)=\sum C(x-y)\mathcal O(\si)$.

\subsection{$J J$}

\paragraph{$J_0J_2$}
\be
\label{ope-j0j2}
           && J_+^{[\mu\nu]}(x) J_-^\mu(y)=\\ \nn
&&          = f^{\mu[\mu\nu]}_\nu\big(\frac{J_+^\nu}{\bar v}+\frac{v}{2\bar v}\p J_+^\nu+\half \pbar J_+^\nu\big)
            + f^{[\mu\nu]}_{\al\ald}f^{\ald \mu}_\bt
                  \big( J_+^\al J_+^\bt\frac{v}{\bar v}
                      - J_+^\al J_-^\bt\log\gm|v|^2\big) \\ \nn
&&          + f^{[\mu\nu]}_{\nu \lm}f^{\mu \lm}_{[\lm\rho]}J_+^\nu\big(N_+^{[\lm\rho]}\frac{v}{\bar v}+N_-^{[\lm\rho]}\log\gm|v|^2\big)
\ee

\be
\label{ope-jbar0j2}
&& J_-^{[\mu\nu]}(x) J_+^\mu(y)=\\ \nn
&&        =  f^{\mu[\mu\nu]}_\nu \big(\frac{J_-^\nu}{v}+
                             \half \p J_-^\nu +
                             \frac{\vbar}{2v}\pbar J_-^\nu\big)
            +f^{[\mu\nu]}_{\al\ald}f^{\al \mu}_{\btd}\big(J_-^{\ald}J_-^{\btd}\frac{\vbar}{v}
                                                   -J_-^{\ald}J_+^{\btd}\log\gm|v|^2\big)\\ \nn
&&          -f^{[\mu\nu]}_{\nu\lm}f^{\lm\mu}_{[\lm\rho]}J_-^\nu\big(N_+^{[\lm\rho]}\log\gm|v|^2+ N_-^{[\lm\rho]}\frac{\vbar}{v}\big)
\ee


\paragraph{$J_0J_3$}
\be
\label{ope-j0jbar3}
&& J_+^{[\mu\nu]}(x)J_-^\al(y)=\\ \nn
&& =  f^{\al[\mu\nu]}_{\bt}\big(\frac{J_+^\bt}{\vbar}+\frac{v}{2\vbar}\p J_+^\bt+\half\pbar J_+^\bt\big)
 + f^{\ald\al}_{[\lm\rho]}f^{[\mu\nu]}_{\bt\ald}J_+^\bt\big(N_-^{[\lm\rho]}\log \gm|v|^2 +N_+^{[\lm\rho]}\frac{v}{\bar v}\big)
 \\ \nn &&
\ee

\be
\label{ope-j3jbar0}
&& J_+^\al(x)J_-^{[\mu\nu]}(y)=\\ \nn
&&    f^{[\mu\nu]\al}_{\bt}\big(\frac{J_-^\bt}{v}
                                -\half\p J_-^\bt
                                -\frac{\vbar}{2v}\pbar J_-^\bt\big)\\ \nn
&&   +f^{[\mu\nu]}_{\mu\nu}f^{\nu\al}_{\btd}\big(J_-^\mu J_-^{\btd}\frac{\vbar}{v}
                                   -J_+^{\btd} J_-^\mu\log\gm|v|^2\big)
     +f^{[\mu\nu]}_{\btd\gm}f^{\gm\al}_\mu\big(J_+^\mu J_-^{\btd}\log\gm|v|^2
                                       -J_-^{\btd} J_-^\mu\frac{\vbar}{v}\big)\\ \nn
&&   +f^{[\mu\nu]}_{\bt\gmd}f^{\gmd\al}_{[\lm\rho]}J_-^\bt\big(N^{[\lm\rho]}_+\log\gm|v|^2+N_-^{[\lm\rho]}\frac{\vbar}{v}\big)
\ee

\paragraph{$J_0 J_1$}

\be
\label{ope-jbar0j1}
&&  J_-^{[\mu\nu]}(x)J_+^{\ald}(y)=\\ \nn
&&  = f^{\ald [\mu\nu]}_{\btd}\big(J_-^{\btd}\frac{1}{v}+
                           \half \p J_-^{\btd}+
                           \frac{\vbar}{2v}\pbar J_-^{\btd}\big)+
  f^{[\mu\nu]}_{\al\btd}f^{\al\ald}_{[\lm\rho]}\big( J_-^{\btd}N^{[\lm\rho]} \log\gm|v|^2+
                                            J_-^{\btd} N_-^{[\lm\rho]} \frac{\vbar}{v}  \big)\\ \nn &&
\ee
\be
\label{ope-j0jbar1}
&&  J_+^{[\mu\nu]}(x)J_-^{\ald}(y)=\\ \nn
&& = f^{\ald [\mu\nu]}_{\btd}\big(\frac{J_+^{\btd}}{\vbar}+
                           \frac{v}{2\vbar}\p J_+^{\btd}+
                           \half \pbar J_+^{\btd}\big)
  + f^{[\mu\nu]}_{\mu\nu}f^{\nu\ald}_\bt \big(J_+^\mu J_+^{\bt}\frac{v}{\vbar}
                                    -J_+^\mu J_-^{\bt}\log\gm|v|^2\big)\\ \nn
&&  + f^{[\mu\nu]}_{\bt\gmd}f^{\ald\gmd}_\mu \big(-J_+^\mu J_+^{\bt}\frac{v}{\vbar}
                                          +J_+^\bt J_-^\mu\log\gm|v|^2\big)
  + f^{[\mu\nu]}_{\al\btd}f^{\al\ald}_{[\lm\rho]}J_+^{\btd}\big(N_+^{[\lm\rho]}\frac{v}{\vbar}+ N_-^{[\lm\rho]}\log\gm|v|^2\big)\\ \nn &&
\ee


\paragraph{$J_0 J_0$}

\be
\label{ope-j0jbar0}
&&  J_+^{[\mu_1\nu_1]}(x)J_-^{[\mu_2\nu_2]}(y) =\\ \nn
&& = \big( f^{[\mu_1\nu_1]\lm}_\mu f^{[\mu_2\nu_2]}_{\nu\lm}J^\mu_+ J_-^\nu
  +f^{\bt[\mu_1\nu_1]}_\al f^{[\mu_2\nu_2]}_{\bt\btd}J_+^\al J_-^{\btd}
  +f^{\btd[\mu_1\nu_1]}_{\ald} f^{[\mu_2\nu_2]}_{\bt\btd}J_+^{\ald}J_-^{\bt}\big)\log\gm|v|^2\\ \nn &&
\ee


\paragraph{$J_3 J_3$}

\be
\label{opej-j3j3}
&& J_+^\al(x)J_-^\bt(y)=\cr
&&  f^{\al\bt}_\mu \frac{J_+^\mu}{\bar v}
    + f^{\al\bt}_\mu \big(\frac{v}{2\bar v} \p J^\mu_+ +\half \pbar J_+^\mu-\half \p J_-^\mu \log\gm|v|^2 \big)\cr
&&  -\half \log\gm|v|^2 \big(f^\bt_{[\mu\nu]\dl}f^{\al[\mu\nu]}_\gm -f^\al_{[\mu\nu]\dl}f^{\bt[\mu\nu]}_\gm\big)J_-^\gm J_+^\dl \cr
&&  +\half \log\gm|v|^2 \big(f^{\al\gm}_\mu f^\bt_{\gm[\mu\nu]}-f^{\bt\gm}_\mu f^\al_{\gm[\mu\nu]}\big)
      \big(N_+^{[\mu\nu]}J_-^\mu+ N_-^{[\mu\nu]}J^\mu_+\big)\cr
&&  -f^{\bt\gm}_\mu f^\al_{\gm[\mu\nu]}J_-^\mu\big(N_+^{[\mu\nu]}\log\gm|v|^2+N_-^{[\mu\nu]}\frac{\bar v}{v}\big)
    +f^\al_{\mu\ald}f^{\ald\bt}_{[\mu\nu]} J_+^\mu\big(N_+^{[\mu\nu]}\frac{v}{\bar v}+N_-^{[\mu\nu]}\log\gm|v|^2\big)  \cr
&&    \ee

\paragraph{$J_1 J_1$}

\be
\label{ope-jbar1j1}
&&  J_-^{\ald}(x)J_+^{\btd}(y)=\\ \nn
&&    = f^{\ald\btd}_\mu\big( \frac{1}{v}J_-^\mu+
                        \half \p J_-^\mu
                        -\half\pbar J_+^\mu \log\gm|v|^2
                        +\frac{\vbar}{2v}\pbar J_-^\mu\big)\\ \nn
&&    +\half \log\gm|v|^2 \big(f^{\ald}_{[\mu\nu]\dld}f^{\btd [\mu\nu]}_{\gmd}-f^{\btd}_{[\mu\nu]\dld}f^{\ald[\mu\nu]}_{\gmd}\big)
      J_-^{\gmd}J_+^{\dld}\\ \nn
&&    +\half \log\gm|v|^2 \big( f^{\ald\gmd}_\mu f^{\btd}_{\gmd[\mu\nu]}-f^{\btd\gmd}_\mu f^{\ald}_{\gmd[\mu\nu]}\big)
                       \big(N^{[\mu\nu]}_+ J_-^\mu + N_-^{[\mu\nu]}J_+^\mu\big)\\ \nn
&&    +f^{\ald}_{\mu\al}f^{\al\btd}_{[\mu\nu]}J_-^\mu\big(N_-^{[\mu\nu]}\frac{\vbar}{v}+N_+^{[\mu\nu]}\log\gm|v|^2\big)
      -f^{\ald}_{\mu\al}f^{\al\btd}_{[\mu\nu]}J_+^\mu\big(N_-^{[\mu\nu]}\log\gm|v|^2+N_+^{[\mu\nu]}\frac{v}{\vbar}\big)\\ \nn &&
\ee

\be
\label{ope-j1jbar1}
&&   J_+^{\ald}(x)J_-^{\btd}(y)=\\ \nn
&&   = f^{\ald\btd}_\mu\big( \frac{J_-^\mu}{v}
                        -\half \p J_-^\mu
                        +\half \pbar J_+^\mu\log\gm|v|^2
                        -\frac{\vbar}{2v}\pbar J_-^\mu\big)\\ \nn
&&   = +\half \log\gm|v|^2 \big(f^{\ald}_{[\mu\nu]\dld}f^{\btd [\mu\nu]}_{\gmd}-f^{\btd}_{[\mu\nu]\dld}f^{\ald[\mu\nu]}_{\gmd}\big)
      J_-^{\gmd}J_+^{\dld}\\ \nn
&&    +\half \log\gm|v|^2 \big( f^{\ald\gmd}_\mu f^{\btd}_{\gmd[\mu\nu]}-f^{\btd\gmd}_\mu f^{\ald}_{\gmd[\mu\nu]}\big)
                       \big(N_+^{[\mu\nu]}J_-^\mu + N_-^{[\mu\nu]}J_+^\mu\big)\\ \nn
&&    -f^{\btd}_{\mu\al}f^{\al\ald}_{[\mu\nu]}J_-^\mu\big(N_-^{[\mu\nu]}\frac{\vbar}{v}+ N_+^{[\mu\nu]}\log\gm|v|^2\big)
      +f^{\btd}_{\mu\al}f^{\al\ald}_{[\mu\nu]}J_+^\mu\big(N_-^{[\mu\nu]}\log\gm|v|^2 +  N_+^{[\mu\nu]}\frac{v}{\vbar}\big)\\ \nn &&
\ee

\paragraph{$J_2J_3$}

\be
\label{ope-j2jbar3}
&& J_+^\mu(x) J_-^\al(y)=\\ \nn
&&=   f^{\al \mu}_{\ald}\big(\frac{J_+^{\ald}}{\vbar}
                            +\frac{v}{2\vbar}\p J_+^{\ald}
                            +\half\pbar J_+^{\ald}
                            -\half \p J_-^{\ald}\log\gm|v|^2\big)\\ \nn
&&    +f^{\mu\gm}_{\ald}f^\al_{\gm[\mu\nu]}J_-^{\ald}\big(N^{[\mu\nu]}_+\log\gm|v|^2+ N_-^{[\mu\nu]}\frac{\vbar}{v}\big)
      -f^{\mu\gm}_{\ald}f^\al_{\gm[\mu\nu]} J_+^{\ald}\big(N^{[\mu\nu]}_+\frac{v}{\vbar}+ N_-^{[\mu\nu]}\log\gm|v|^2\big) \\ \nn
&&     + \textbf{R}^{\mu\al}_{+-}\log\gm|v|^2
\ee

\be
\label{ope-j3jbar2}
&& J_+^{\al}(x)J_-^\mu(y)=\\ \nn
&&   f^{\mu\al}_{\ald}\big(\frac{J_+^{\ald}}{\vbar}
                      +\frac{v}{2\vbar}\p J_+^{\ald}
                      +\half\pbar J_+^{\ald}
                      -\half \p J_-^{\ald}\log\gm|v|^2\big)\\ \nn
&&    +f^{\al\gm}_\nu f^\mu_{\gm\bt}\big(J_-^\nu J_+^\bt \log\gm|v|^2
                                -J_+^\nu J_+^\bt\frac{v}{\vbar}
                                -J_-^\nu J_-^\bt\frac{\vbar}{v}
                                +J_+^\nu J_-^\bt \log\gm|v|^2\big)\\ \nn
&&    +f^{\al \nu}_{\ald}f^\mu_{\nu[\mu\nu]} J_-^{\ald}\big(N_+^{[\mu\nu]}\log\gm|v|^2+  N_-^{[\mu\nu]}\frac{\vbar}{v}\big)
      -f^{\al \nu}_{\ald}f^\mu_{\nu[\mu\nu]} J_+^{\ald}\big(N_+^{[\mu\nu]}\frac{v}{\vbar}+  N_-^{[\mu\nu]}\log\gm|v|^2\big)\\ \nn
&&     +\textbf{R}^{\al\mu}_{+-}\log\gm|v|^2
\ee
The tensor $\textbf{R}^{\mu\al}_{+-}$ is a symmetric tensor and it contains all the terms coming from the diagram computed from the vertices \eqref{double-ins-mg} and \eqref{double-ins-mm}. They diverge logarithmically however these type of insertions being symmetric are just cancelled when we take the sum of the commutator between $J_+^\mu(x) J_-^\al(y)$ and $J_+^{\al}(x)J_-^\mu(y)$.

\paragraph{$J_1 J_2$}
The same structure as before for the case $J_2 J_3$ appears here.

\be
\label{ope-j1jbar2}
&&  J_+^{\ald}(x)J_-^\mu(y)=\\ \nn
&&     = f^{\ald \mu}_\bt \big(-\frac{J_-^\bt}{v}
                      -\half \pbar J_+^\bt\log\gm|v|^2
                      +\half \frac{\vbar}{v}\pbar J_-^\bt
                      +\half \p J_-^\bt\big)\\ \nn
&&     +f^\mu_{\bt\gm}f^{\gm\ald}_{[\mu\nu]} J_-^\bt\big(N_+^{[\mu\nu]}\log\gm|v|^2 +  N_-^{[\mu\nu]}\frac{\vbar}{v}\big)
       +f^{\ald}_{[\mu\nu]\gmd}f^{\gmd \mu}_\bt J_+^\bt\big(N_+^{[\mu\nu]}\frac{v}{\vbar}+  N_-^{[\mu\nu]}\log\gm|v|^2\big)\\ \nn
&&     +\textbf{R}^{\ald\mu}_{+-}\log\gm|v|^2
\ee

\be
\label{ope-j2jbar1}
&& J^\mu_+(x)J_-^{\ald}(y)=\\ \nn
&&       =f^{ \mu\ald}_\bt \big(-\frac{J_-^\bt}{v}
                               +\half \frac{\vbar}{v}\pbar J_-^\bt
                               +\half \p J_-^\bt
                              -\half \pbar J_+^\bt\log\gm|v|^2\big)\\ \nn
&&      +f^\mu_{\gmd\btd}f^{\ald\gmd}_\nu \big(J_-^{\btd}J_+^\nu \log\gm|v|^2
                                    -J_-^\nu J_-^{\btd}\frac{\vbar}{v}
                                    -J_+^{\btd} J_-^\nu\frac{v}{\vbar}
                                    +J_-^\nu J_+^{\btd}\log\gm|v|^2\big)\\ \nn
&&       +f^{\mu\nu}_{[\mu\nu]}f^{\ald}_{\nu\bt}J_+^\bt\big(N_+^{[\mu\nu]}\frac{v}{\vbar}+ N_-^{[\mu\nu]}\log\gm|v|^2\big)
        +f^{\mu\nu}_{[\mu\nu]}f^{\ald}_{\nu\bt}J_-^{\bt}\big(N_+^{[\mu\nu]}\log\gm|v|^2 + N_-^{[\mu\nu]}\frac{\vbar}{v}\big)\\ \nn
&&         +\textbf{R}^{\ald\mu}_{+-}\log\gm|v|^2
\ee
Again $\textbf{R}^{\ald\mu}_{+-}$ is the same kind of tensor as before, it comes from the same vertices \eqref{double-ins-mg} and \eqref{double-ins-mm}, with the replacement $\al\rightarrow\ald$.

\paragraph{$J_2J_2$}

\be
\label{ope-j2j2bar}
&& J^\mu_+(x)J_-^\nu(y)=\\ \nn
&&       = -f^{\mu\nu}_{[\mu\nu]}\big(\frac{N_+^{[\mu\nu]}}{\vbar}+\frac{N_-^{[\mu\nu]}}{v}\big)\\ \nn
&&         +\half f^{\mu\nu}_{[\mu\nu]}\big( -\frac{v}{\vbar}\p N^{[\mu\nu]}_++\pbar N^{[\mu\nu]}_+(-1+\log\gm|v|^2)
                                             +\frac{\vbar}{v}\pbar N^{[\mu\nu]}_-+\p N^{[\mu\nu]}_-(1-\log\gm|v|^2)\big)+\\ \nn
&&        -f^{\mu\lm}_{[\mu_1\nu_1]}f^\nu_{\lm[\mu_2\nu_2]}\big(N^{[\mu_1\nu_1]}_+N^{[\mu_2\nu_2]}_+\frac{v}{\vbar}
                                                                + N^{[\mu_1\nu_1]}_-N^{[\mu_2\nu_2]}_-\frac{\vbar}{v}\big)\\ \nn
&&        -f^{\mu\lm}_{[\mu_1\nu_1]}f^\nu_{\lm[\mu_2\nu_2]}\big( N^{[\mu_1\nu_1]}_+N^{[\mu_2\nu_2]}_-\log\gm|v|^2
                                                                + N^{[\mu_1\nu_1]}_-N^{[\mu_2\nu_2]}_+\log\gm|v|^2\big)\\ \nn
&&        +f^{\gm\mu}_{\btd}f^\nu_{\gm\al}\big(J_-^{\btd}J_+^{\al}+J^{\btd}_+J^\al_-\big) \log\gm|v|^2
          +f^\mu_{\ald\btd}f^{\btd\nu}_\bt J^{\ald}_+J^\bt_+\frac{v}{\vbar}
          +f^\nu_{\al\bt}f^{\bt\mu}_{\ald}J^\al_- J^{\ald}_- \frac{\vbar}{v}\\ \nn
&&        -\half \big(f^{\mu[\mu\nu]}_\lm f^\nu_{[\mu\nu]\rho}+ f^{\nu[\mu\nu]}_\lm f^\mu_{[\mu\nu]\rho}\big) J^\lm_-J^\rho_+\log\gm|v|^2\\ \nn
&&        -\half\big(f^{\mu\gmd}_\al f^\nu_{\gmd\btd}+ f^{\nu\gmd}_\al f^\mu_{\gmd\btd}\big)J^\al_- J^{\btd}_+ \log\gm|v|^2
          +\half\big(f^{\mu\gm}_{\btd} f^\nu_{\gm\al}+ f^{\nu\gm}_{\btd} f^\mu_{\gm\al}\big) J^{\btd}_-J^\al_+ \log\gm|v|^2
\ee


\paragraph{$J_1J_3$}

\be
\label{ope-j1j3bar}
&& J^{\ald}_+(x) J^\bt_-(y)=\\ \nn
&&     = f^{\ald\bt}_{[\mu\nu]}\big(\frac{N_+^{[\mu\nu]}}{\vbar}+\frac{N_-^{[\mu\nu]}}{v}\big)\\ \nn
&&      +\half f^{\ald\bt}_{[\mu\nu]}\big( \frac{v}{\vbar}\p N^{[\mu\nu]}_+ +\pbar N^{[\mu\nu]}_+(1-\log\gm|v|^2)
                                           -\frac{\vbar}{v}\pbar N^{[\mu\nu]}_- -\p N^{[\mu\nu]}_-(1-\log\gm|v|^2)\big)\\ \nn
&&      +f^{\ald\gm}_{[\mu_1\nu_1]}f^{\bt}_{\gm[\mu_2\nu_2]}\big(N^{[\mu_1\nu_1]}_+N^{[\mu_2\nu_2]}_+\frac{v}{\vbar}
                                                                + N^{[\mu_1\nu_1]}_-N^{[\mu_2\nu_2]}_-\frac{\vbar}{v}\big)\\ \nn
&&      +f^{\ald\gm}_{[\mu_1\nu_1]}f^{\bt}_{\gm[\mu_2\nu_2]}\big( N^{[\mu_1\nu_1]}_+N^{[\mu_2\nu_2]}_-\log\gm|v|^2
                                                                + N^{[\mu_1\nu_1]}_-N^{[\mu_2\nu_2]}_+\log\gm|v|^2\big)\\ \nn
&&      +\frac{1}{4}\big(3 f^{\ald}_{[\mu\nu]\btd}f^{\bt[\mu\nu]}_\al -f^\bt_{\mu\btd}f^{\ald\mu}_\al\big)
                     J^\al_- J^{\btd}_+\log\gm|v|^2
       +\frac{1}{4}f^{[\mu\nu]}_{\al\btd}f^{\ald\bt}_{[\mu\nu]} J^{\btd}_- J^\al_+\log\gm|v|^2\\ \nn
&&        +\frac{1}{4}f^{\ald\bt}_{[\mu\nu]}f^{[\mu\nu]}_{\mu\nu}J^\mu_- J^\nu_+ \log\gm|v|^2
\ee

\paragraph{$J_3 J_1$}

\be
\label{ope-j3j1bar}
&& J^\bt_+(x) J^{\ald}_-(y)=\\ \nn
&&       = f^{\ald\bt}_{[\mu\nu]}\big(\frac{N_+^{[\mu\nu]}}{\vbar}+\frac{N_-^{[\mu\nu]}}{v}\big)\\ \nn
&&       +\half f^{\ald\bt}_{[\mu\nu]}\big( \frac{v}{\vbar}\p N^{[\mu\nu]}_+ +\pbar N^{[\mu\nu]}_+(1-\log\gm|v|^2)
                                           -\frac{\vbar}{v}\pbar N^{[\mu\nu]}_- -\p N^{[\mu\nu]}_-(1-\log\gm|v|^2)\big)\\ \nn
&&      +f^{\bt\gmd}_{[\mu_1\nu_1]}f^{\ald}_{\gmd[\mu_2\nu_2]}\big(N^{[\mu_1\nu_1]}_+N^{[\mu_2\nu_2]}_+\frac{v}{\vbar}
                                                                + N^{[\mu_1\nu_1]}_-N^{[\mu_2\nu_2]}_-\frac{\vbar}{v}\big)\\ \nn
&&      +\big(f^{\bt\gm}_\mu f^{\ald}_{\gm\nu}-f^{\ald}_{\mu\gm}f^{\gm\bt}_\nu\big)J^\mu_- J^\nu_+ \log\gm|v|^2
        -f^\bt_{\mu\gmd}f^{\gmd\ald}_\nu J^\mu_+ J^\nu_+ \frac{v}{\vbar}
        -f^{\ald}_{\mu\gm}f^{\bt\gm}_\nu J^\mu_- J^\nu_- \frac{\vbar}{v}\\ \nn
&&      -f^{\bt\mu}_{\btd}f^{\ald}_{\mu\al}J_-^{\btd}J^\al_+ \log\gm|v|^2
        +f^\bt_{\btd \mu}f^{\mu\ald}_\al J^{\btd}_+ J^\al_+ \frac{v}{\vbar}
        +f^{\ald}_{\al\mu}f^{\bt\mu}_{\btd}J^\al_- J^{\btd}_-\frac{\vbar}{v}
        +f^{\ald}_{\al\mu}f^{\bt\mu}_{\btd}J^{\btd}_+ J^\al_- \log\gm|v|^2\\ \nn
&&      +\frac{1}{4}\big( 3f^{\ald}_{[\mu\nu]\btd}f^{[\mu\nu]\bt}_\al -f^\bt_{\mu\btd}f^{\mu\ald}_{[\mu\nu]}\big)
                         J^\al_-J^{\btd}_+ \log\gm|v|^2\\ \nn
&&     -\frac{1}{4}f^{\ald\bt}_{[\mu\nu]}f^{[\mu\nu]}_{\mu\nu}J^\mu_- J^\nu_+ \log\gm|v|^2
       -\frac{1}{4}f^{[\mu\nu]}_{\al\btd}f^{\ald\bt}_{[\mu\nu]}J^{\btd}_- J^{\al}_+ \log\gm|v|^2
\ee


\subsection{$J N$}

\be
\label{ope-j2nbar}
 J^\mu_+(x)N_-^{[\mu\nu]}(y)=
      f^\mu_{\nu[\mu_1\nu_1]}f^{[\mu\nu][\mu_1\nu_1]}_{[\mu_2\nu_2]}N_-^{[\mu_2\nu_2]}J_+^\nu\log\gm|v|^2
\ee
\be
\label{ope-j2barn}
J_-^\mu(x)N_+^{[\mu\nu]}(y)=
     f^\mu_{\nu[\mu_1\nu_1]}f^{[\mu\nu][\mu_1\nu_1]}_{[\mu_2\nu_2]}N_+^{[\mu_2\nu_2]}J_-^\nu\log\gm|v|^2
\ee
\be
\label{ope-j3nbar}
J_+^\al(x)N_-^{[\mu\nu]}(x)=
 f^\al_{[\mu_1\nu_1]\bt}f^{[\mu_1\nu_1][\mu\nu]}_{[\mu_2\nu_2]}J_+^\bt N_-^{[\mu_2\nu_2]}\log\gm|v|^2
\ee
\be
\label{ope-jbar3n}
J_-^\al(x)N_+^{[\mu\nu]}(y)=
f^\al_{[\mu_1\nu_1]\bt}f^{[\mu_1\nu_1][\mu\nu]}_{[\mu_2\nu_2]}J_-^\bt N_+^{[\mu_2\nu_2]}\log\gm|v|^2
\ee
\be
\label{ope-jbar1n}
J_-^{\ald}(x)N_+^{[\mu\nu]}(y)=
 f^{[\mu\nu][\mu_1\nu_1]}_{[\mu_2\nu_2]}f^{\ald}_{\btd [\mu_1\nu_1]}J_-^{\btd} N^{[\mu_2\nu_2]}_+\log\gm|v|^2
\ee
\be
\label{ope-j1nbar}
J_+^{\ald}(x)N_-^{[\mu\nu]}(y)=
 f^{[\mu\nu][\mu_1\nu_1]}_{[\mu_2\nu_2]}f^{\ald}_{\btd [\mu_1\nu_1]}J_+^{\btd} N^{[\mu_2\nu_2]}_-\log\gm|v|^2
\ee

\subsection{$ N N$}
\be
N^{[\mu\nu]}_{-}(x)N^{[\lm\rho]}_{+}(y)
= f^{[\mu\nu]}_{[\mu_1\nu_1][\mu_2\nu_2]}f^{[\lm\rho][\mu_2\nu_2]}_{[\mu_3\nu_3]}N^{[\mu_3\nu_3]}_+ N_-^{[\mu_1\nu_2]}\log\gm|v|^2
\ee


\section{Useful algebraic identities}
\label{app:alg-identities}

Here we list the algebraic identities abundantly used in the calculations. Some of them were derived in \cite{Mikhailov:2007mr}.
\[
\label{id1}
f_{\al\mu}^{\ald}f_{\btd}^{\al\mu}
= f_{\ald\mu}^{\al}f_{\bt}^{\ald\mu}
= f_{\gmd[\mu\nu]}^{\ald}f_{\btd}^{\gmd[\mu\nu]}
= f_{\gm[\mu\nu]}^{\al}f_{\bt}^{\gm[\mu\nu]}=0
\]

\[
\label{id2}
2 f^{[\mu_1\nu_1]}_{\gmd\gm} f^{\gm\gmd}_{[\mu_2\nu_2]}
+ f^{[\mu_1\nu_1]}_{\rho_1\rho_2} f^{\rho_2\rho_1}_{[\mu_2\nu_2]}
+f^{[\mu_1\nu_1]}_{[\mu_3\nu_3][\mu_4\nu_4]}f^{[\mu_4\nu_4][\mu_3\nu_3]}_{[\mu_2\nu_2]}=0
\]

\[
\label{id3}
f^\mu_{\btd\gmd}f^{\btd\gmd}_\nu=
f^\mu_{\bt\gm}f^{\bt\gm}_\nu=
f^\mu_{\lm_1\rho_1}f^{\rho_1\lm_1}_\nu
\]

\[
\label{id4}
f^{[\mu\nu]}_{\gmd\gm}C^{\gmd\gm}=f^{[\mu\nu]}_{\mu\nu} C^{\mu\nu}=0
\]

\[
\label{id5}
f^\lm_{\mu[\mu\nu]}f^{\al\bt}_\lm f^{\nu}_{\al\bt}=
f^\nu_{\lm[\mu\nu]}f^\lm_{\al\bt} f^{\al\bt}_\mu
\]

\be
\label{id6}
f^{[\mu_1\nu_1][\mu_2\nu_2]}_{[\mu_4\nu_4]}f^{[\mu_3\nu_3]}_{[\mu_2\nu_2][\mu_1\nu_1]}f^{[\mu_4\nu_4]}_{[\mu_5\nu_5][\mu_6\nu_6]}=
f^{[\mu_1\nu_1][\mu_2\nu_2]}_{[\mu_5\nu_5]}f^{[\mu_4\nu_4]}_{[\mu_2\nu_2][\mu_1\nu_1]}f^{[\mu_3\nu_3]}_{[\mu_4\nu_4][\mu_6\nu_6]}=\cr
=f^{[\mu_1\nu_1][\mu_2\nu_2]}_{[\mu_6\nu_6]}f^{[\mu_4\nu_4]}_{[\mu_2\nu_2][\mu_1\nu_1]}f^{[\mu_3\nu_3]}_{[\mu_5\nu_5][\mu_4\nu_4]}
\ee


\section{Field strength: some further results}
\label{app:comm-results}

In this section we write further examples in order to compute the field strength. In particular the sectors labelled by $z^{3/2}\;,z^{0}$ and $z^2$. We stress ones more that thanks to the symmetry \eqref{symm-f} all the remaining sectors can be obtained from those reported in the paper.


\subsection{Commutators of $z^{3/2}$ }

We rewrite here the commutators to compute for the case $z^{3/2}$
\[
\label{mu3-2}
[J_{1+},\Jom]-[J_{1+},\Nm]+[\Np,\Jonem]
\]


\paragraph{OPE's.}

\begin{itemize}
\item The derivatives, which are contained only in $[J_{1+},\Jom]$ \eqref{ope-jbar0j1}, do not contribute due to the algebraic identities $f^{\gmd}_{[\mu\nu]\ald} f^{\ald[\mu\nu]}_{\btd}=0$ \eqref{id1}.
\item The only possible contributions are the mixed commutators. From $[J_{1+},\Nm]$, $[\Np,\Jonem]$, \eqref{ope-jbar1n} and \eqref{ope-j1nbar} one has
\be
- f^{\gmd}_{[\mu_1\nu_1]\ald} f^{\ald}_{\btd[\mu_2\nu_2]}f^{[\mu_2\nu_2][\mu_1\nu_1]}_{[\mu_3\nu_3]}
  \big(J_+^{\btd}N_-^{[\mu_3\nu_3]}+J_-^{\btd}N_+^{[\mu_3\nu_3]}\big)\log\gm|v|^2 = \cr
-\half f^{[\mu_2\nu_2][\mu_1\nu_1]}_{[\mu_4\nu_4]}f_{[\mu_1\nu_1][\mu_2\nu_2]}^{[\mu_3\nu_3]}f^{\gmd}_{\btd [\mu_3\nu_3]}
  \big(J_+^{\btd}N_-^{[\mu_4\nu_4]}+J_-^{\btd}N_+^{[\mu_4\nu_4]}\big)\log\gm|v|^2\;,
\ee
while $[J_{1+},\Jom]$ \eqref{ope-jbar0j1} gives
\be
&&- f^{\gmd}_{[\mu_1\nu_1]\ald} f^{[\mu_1\nu_1]}_{\al\btd}f^{\al\ald}_{[\mu_2\nu_2]}
    J_-^{\btd} N_+^{[\mu_2\nu_2]}\log\gm|v|^2=\\ \nn %
&&= \half f^{[\mu_2\nu_2][\mu_1\nu_1]}_{[\mu_4\nu_4]}f_{[\mu_1\nu_1][\mu_2\nu_2]}^{[\mu_3\nu_3]}f^{\gmd}_{\btd [\mu_3\nu_3]}
    J_-^{\btd} N_+^{[\mu_4\nu_4]}\log\gm|v|^2 +...\;,
\ee
with the help of the identities in \Appref{app:alg-identities} and the Jacobi identities.
\end{itemize}
Summing the two contributions above the result is
\be
\label{result-comm-mu3-2}
-\half f^{[\mu_2\nu_2][\mu_1\nu_1]}_{[\mu_2\nu_2]}f_{[\mu_1\nu_1][\mu_2\nu_2]}^{[\mu_3\nu_3]}f^{\gmd}_{\btd[\mu_3\nu_3]}
       J_+^{\btd}N_-^{[\mu_2\nu_2]}\log\gm|v|^2
\ee


\paragraph{Internal contractions.}
Looking at the expansion for the commutators $-[J_{1+},\Nm]$ and $[\Np,\Jonem]$ one sees directly that they do not contain logarithmic terms, since again bosonic and fermionic indices are contracted in two summed structure constants \eqref{id1}.
From the internal contractions of $J_0$ the commutator $[J_{1+},\Jom]$ produces
\be
:[J_{1+},\Jom]:\,=
\half [J_{1+},\langle [\pbar X_3,X_1]+[\pbar X_1,X_3]+[\pbar X_2,X_2]\rangle ]+...=\cr
= \half f^{\gmd}_{\btd[\mu_1\nu_1]}f^{[\mu_1\nu_1]}_{[\mu_3\nu_3][\mu_4\nu_4]}f^{[\mu_4\nu_4][\mu_3\nu_3]}_{[\mu_2\nu_2]}
      J_+^{\btd}N_-^{[\mu_2\nu_2]}\log\gm|v|^2+...\;.
\ee
which cancels the divergence computed in \eqref{result-comm-mu3-2}.


\subsection{$z^0$}

The expression to verify for the sector labelled by $z^0$ is
\[
\label{z0-comm}
[\Jop,\Jom]-[\Jom,\Nm]-[\Np,\Jom] + 2[\Np,\Nm]
    +[\Jtwop,\Jtwom]+[\Jthreep,\Jonem]+[\Jonep,\Jthreem]\;.
\]
This case is slightly different from all the others. In fact the OPE contributions to \eqref{z0-comm} are finite by themselves, as one sees in the next paragraph, but the internal contractions for the commutators are logarithmically divergent. Consequently the terms in $\p\Jflat_--\pbar\Jflat_+$ \eqref{F-derivatives} are important in order to cancel such divergences and leave all the $z^0$ sector finite. This is the only case where the terms \eqref{F-derivatives} contribute.

\paragraph{OPE's.}

The commutators $[J_{0+},\Nm],\;[\Np,\Jom]$ do not produce any term of dimension two at this order.

\begin{itemize}
\item For the derivatives only the matter commutators $[\Jtwop,\Jtwom]$, $[\Jonep,\Jthreem]$ and $[\Jthreep,\Jonem]$  contribute with the OPE's \eqref{ope-j1j3bar}, \eqref{ope-j2j2bar}, \eqref{ope-j3j1bar}:
\be
 -f^{[\mu\nu]}_{[\mu_1\nu_1][\mu_2\nu_2]}f^{[\mu_1\nu_1][\mu_2\nu_2]}_{[\mu_3\nu_3]}
           \big( -\half \pbar N^{[\mu_3\nu_3]}_+ 
                  \log\gm|v|^2
              +\half \p N^{[\mu_3\nu_3]}_- 
              log\gm|v|^2
              \big)\;.
\ee
\item For the commutators of matter currents the contributions come from all the commutators. Using the Jacobi identities one can see that all the terms coming from $[\Jonep,\Jthreem]$ \eqref{ope-j1j3bar}, $[\Jtwop,\Jtwom]$ \eqref{ope-j2j2bar}, $[\Jthreep,\Jonem]$ \eqref{ope-j3j1bar} cancel. Thus the only contribution which remains is from the OPE $[\Jop,\Jom]$ \eqref{ope-j0jbar0}, i.e.
\be
\half f^{[\mu\nu]}_{[\mu_1\nu_1][\mu_2\nu_2]}f^{[\mu_1\nu_1][\mu_2\nu_2]}_{[\mu_3\nu_3]}
      \big(f^{[\mu_3\nu_3]}_{\mu\nu}J^\mu_+ J_-^\nu
          +f^{[\mu_3\nu_3]}_{\al\btd} J^\al_+ J^{\btd}_-
          +f^{[\mu_3\nu_3]}_{\ald\bt}J^{\ald}_+ J^\bt_-\big)\log\gm|v|^2
\ee
\item Finally for the commutators of ghost currents the only OPE which does not contribute is $[J_{0+},J_{0-}]$ \eqref{ope-j0jbar0}. Summing $[\Jonep,\Jthreem]$ \eqref{ope-j1j3bar}, $[\Jtwop,\Jtwom]$ \eqref{ope-j2j2bar}, $[\Jthreep,\Jonem]$ \eqref{ope-j3j1bar} one has
\be
\half f^{[\mu\nu]}_{[\mu_1\nu_1][\mu_2\nu_2]}f^{[\mu_2\nu_2][\mu_1\nu_1]}_{[\mu_3\nu_3]}
      f^{[\mu_3\nu_3]}_{[\lm_1\rho_1][\lm_2\rho_2]}\big(
     N^{[\lm_1\rho_1]}_- N^{[\lm_2\rho_2]}_+ + N^{[\lm_1\rho_1]}_+ N^{[\lm_2\rho_2]}_-\big)\log\gm|v|^2\;,
\ee
while from $[\Np,\Nm]$ \eqref{ope-nn} one gets
\be
f^{[\mu\nu]}_{[\mu_1\nu_1][\mu_2\nu_2]}f^{[\mu_2\nu_2][\mu_1\nu_1]}_{[\mu_3\nu_3]}
      f^{[\mu_3\nu_3]}_{[\lm_1\rho_1][\lm_2\rho_2]} N^{[\lm_1\rho_1]}_+ N^{[\lm_2\rho_2]}_- \log\gm|v|^2\;.
\ee
Thus the two terms above give
\be
 \half f^{[\mu\nu]}_{[\mu_1\nu_1][\mu_2\nu_2]}f^{[\mu_2\nu_2][\mu_1\nu_1]}_{[\mu_3\nu_3]}
      f^{[\mu_3\nu_3]}_{[\lm_1\rho_1][\lm_2\rho_2]}\big(
      N^{[\lm_1\rho_1]}_+ N^{[\lm_2\rho_2]}_- -
           N^{[\lm_1\rho_1]}_- N^{[\lm_2\rho_2]}_+ \big)\log\gm|v|^2
\;.
\ee
\end{itemize}

Collecting all the different contributions one obtains the following
\be
&&  \half f^{[\mu\nu]}_{[\mu_1\nu_1][\mu_2\nu_2]}f^{[\mu_2\nu_2][\mu_1\nu_1]}_{[\mu_3\nu_3]}
          \big(
               \pbar N^{[\mu_3\nu_3]}_+ -\p N^{[\mu_3\nu_3]}_-
               + f^{[\mu_3\nu_3]}_{[\lm_1\rho_1][\lm_2\rho_2]}N^{[\lm_1\rho_1]}_+ N^{[\lm_2\rho_2]}_-\\ \nn
&&             - f^{[\mu_3\nu_3]}_{[\lm_1\rho_1][\lm_2\rho_2]}N^{[\lm_1\rho_1]}_- N^{[\lm_2\rho_2]}_+
               + f^{[\mu_3\nu_3]}_{\ald\bt}J^{\ald}_+ J^\bt_-
               + f^{[\mu_3\nu_3]}_{\al\btd}J^\al_+ J^{\btd}_-
               + f^{[\mu_3\nu_3]}_{\mu\nu}J^\mu_+ J^\nu_-
               \big)\log\gm|v|^2.
\ee
The first and the second lines are zero due to the classical equations of motion \eqref{eom-matter} and \eqref{eom-ghost}%
\footnote{
In our gauge where $[\widetilde J_{0\pm},X_i]=0$
the Maurer-Cartan identity on the gauge field becomes
\[
[J_{1+},\Jthreem]+ [J_{3+},\Jonem]+ [J_{2+},\Jtwom]=0
\]
which we can consider as equations of motion for the gauge field $J_0$.}.
%
Notice that in this specific case $z^0$ there is \emph{no} logarithmic divergence left.


\paragraph{Internal contractions.}

However the internal contractions for the commutators \eqref{z0-comm} give logarithmic divergences, consequently we get a special and extra contribution from the internal contraction of \eqref{F-derivatives} in order to have also in this sector an UV-finite expression.

The only normal ordered commutators which contribute are $[J_{0+},\Nm]$ and $[\Np,J_{0-}]$, indeed
\be
\label{mu0-comm-no}
&&- :[J_{0+},\Nm]:\,=
  -\half f^{[\mu\nu]}_{[\mu_1\nu_1][\mu_2\nu_2]}f^{[\mu_1\nu_1]}_{[\mu_3\nu_3][\mu_4\nu_4]}
       f^{[\mu_4\nu_4][\mu_3\nu_3]}_{[\lm\rho]} N^{[\mu_2\nu_2]}_-
        N^{[\lm\rho]}_+ \log\gm|v|^2+...
        \\ \nn
&&
-:[\Np,\Jom]:\,=
      -\half f^{[\mu\nu]}_{[\mu_1\nu_1][\mu_2\nu_2]}f^{[\mu_2\nu_2]}_{[\mu_3\nu_3][\mu_4\nu_4]}
       f^{[\mu_4\nu_4][\mu_3\nu_3]}_{[\lm\rho]} N^{[\mu_1\nu_1]}_+
       N^{[\lm\rho]}_- \log\gm|v|^2+...
\;.\ee
The internal contractions for the derivatives are in principle
\[
:\,\p\Jom
-\p \Nm
-\pbar J_{0+}
+\pbar \Np\,:
\]
but the only terms which contribute are
\be
\label{mu0-der}
&&:\,\p \Jom^{(2)}\,:\,= \half f^{[\mu\nu]}_{[\mu_1\nu_1][\mu_2\nu_2]}f_{[\lm\rho]}^{[\mu_2\nu_2][\mu_1\nu_1]}
                \p \Nm^{[\lm\rho]} \log\gm|v|^2+...\\ \nn
&&
:\,-\pbar \Jop^{(2)}\,:\,= -\half f^{[\mu\nu]}_{[\mu_1\nu_1][\mu_2\nu_2]}f_{[\lm\rho]}^{[\mu_2\nu_2][\mu_1\nu_1]}
                \pbar\Np^{[\lm\rho]} \log\gm|v|^2+...
\;.\ee
Using the identities in \appref{app:alg-identities} and the ghost equations of motion \eqref{eom-ghost}, the two contributions \eqref{mu0-comm-no}, \eqref{mu0-der} cancel.


\subsection{Commutators of $z^2$}

Recall that for $z^2$ we have only two terms, i.e.
\[
[\Np,\Jom]
-[\Nm,\Jop]\;.
\]

\paragraph{OPE.} The first terms is of order $\mathcal O(J^2)$ and the latter produces
\[
-[\Np,\Nm]=
-\half f^{[\mu\nu]}_{[\mu_1\nu_1][\mu_2\nu_2]}f^{[\mu_2\nu_2][\mu_1\nu_1]}_{[\mu_3\nu_3]}
       f^{[\mu_3\nu_3]}_{[\lm_1\rho_1][\lm_2\rho_2]}N_+^{[\lm_1\rho_1]}\Nm^{[\lm_2\rho_2]}\log\gm |v|^2
\]

\paragraph{Internal contractions.}
The only log-divergent contributions come from
\be
&&:\,[\Np,\Jom]\,:\,
=\half [\Np,\langle [\pbar X_2,X_2]+[\pbar X_1,X_3]+[\pbar X_3,X_1]\rangle]+...\\ \nn
&&=\half f^{[\mu\nu]}_{[\mu_1\nu_1][\mu_2\nu_2]}f^{[\mu_2\nu_2]}_{[\lm_1\rho_1][\lm_2\rho_2]}f^{[\lm_2\rho_2][\lm_1\rho_1]}_{[\lm_3\rho_3]}
       \Np^{[\mu_1\nu_1]}\Nm^{[\lm_3\rho_3]} \log\gm|v|^2
       \;+....
\ee
Thus the two logarithmic terms cancel using the identity \eqref{id6}.


\end{document}